\documentclass[12pt, a4paper]{article}

\pdfoutput=1

\usepackage{amsmath}
\usepackage{amsfonts}
\usepackage{amssymb}
\usepackage{cite}
\usepackage{multicol,multirow}
\usepackage{graphicx, physics, subcaption}
\usepackage{graphics, epsfig}
\usepackage{epsf}
\usepackage{epstopdf}
\usepackage[usenames,dvipsnames]{color}

\usepackage{colortbl}
\definecolor{summersky}{cmyk}{0.71,0.33,0,0.14}
\definecolor{flamingo}{cmyk}{0,0.51,0.71,0.14}
\definecolor{rp}{cmyk}{0.2, 1, 0.6, 0}
\definecolor{pacificblue}{cmyk}{0.95,0.3,0, 0.19}
\definecolor{gray60}{cmyk}{0.4,0.4,0,0.8}

\numberwithin{equation}{section}

\usepackage[colorlinks,bookmarks=false, hypertexnames=true]{hyperref}
\hypersetup{
	breaklinks=true,
	pdfstartview={FitH},    
	colorlinks=true,       
	linkcolor=blue,          
	citecolor=red,        
	filecolor=magenta,      
	urlcolor=red,           
	anchorcolor=green,      
	linktocpage=true
}

\newcommand{\nc}{\newcommand}

\nc{\ba}{\begin{eqnarray}}
\nc{\ea}{\end{eqnarray}}

\nc{\calR}{{\cal{R}}}
\nc{\calP}{{\cal{P}}}
\nc{\cN}{ {\cal{N}} }

\def\bfx{{\bf x}}

\begin{document}

\def\thefootnote{\fnsymbol{footnote}}

\begin{center}

{\bf\large Revisiting  Magnetogenesis during Inflation
}
\\[0.5cm]

{ Alireza Talebian \footnote{talebian@ipm.ir}, 
Amin Nassiri-Rad \footnote{amin.nassiriraad@ipm.ir}, 
Hassan Firouzjahi \footnote{firouz@ipm.ir},
}
\\[0.5cm]
 
 {\small \textit{$^1$School of Astronomy, Institute for Research in Fundamental Sciences (IPM) \\ P.~O.~Box 19395-5531, Tehran, Iran
}}\\

\end{center}

\vspace{.3cm}
\hrule
\begin{abstract}

We revisit the mechanism of primordial magnetogenesis during inflation by taking into account the dynamics of the stochastic noises of the electromagnetic perturbations. We obtain the associated Langevin and Fokker-Planck equations for the electromagnetic fields and solve them analytically. It is shown that while the backreactions of the electric field energy density may spoil inflation too early, but there are regions of parameter space where the usual decaying behavior of the magnetic fields are replaced by a mean-reverting process of stochastic dynamics. As a result, the magnetic fields settle down into an equilibrium state with the amplitude significantly larger than what is obtained in the absence of the stochastic noises. We show that magnetic fields with present time amplitude $\sim 10^{-13}$ Gauss and coherent length ${\rm Mpc}$ can be generated while the backreactions of the electric field perturbations are under control.  

\end{abstract}

\vspace{0.3cm}
\hrule

\newpage

\section{Introduction}
\label{sec:intro}

Magnetic fields are permeated through celestial bodies, from planets and interstellar mediums to galaxies, galactic clusters (with magnetic field $\sim$ micro-Gauss \cite{Bernet:2008qp,Bonafede:2010xg,Feretti:2012vk}) and intergalactic medium (with magnetic field $\sim$ femto-Gauss  ~\cite{Neronov:1900zz,Tavecchio:2010mk,Dermer:2010mm,Vovk:2011aa,Tavecchio:2010ja,Dolag:2010ni,Taylor:2011bn, Takahashi:2011ac,Huan:2011kp,Finke:2015ona}). Particularly interesting cases are magnetic fields with very large coherence length scale $\lambda_{\rm B} \gtrsim 1\,{\rm Mpc}$ detected in cosmic voids. Several studies~\cite{Neronov:1900zz, Dolag:2010ni, Essey:2010nd, Taylor:2011bn, Chen:2014rsa,Finke:2015ona,Biteau:2018tmv, Feretti:2012vk,Tavecchio:2010mk,Caprini:2015gga} have claimed that gamma-ray observations of distant TeV blazars place lower bounds on the magnetic field strength on these very large scales. Combining the corresponding results with the CMB observations, POLARBEAR and NRAO VLA Sky Survey \cite{Ade:2015cva,Ade:2015cao,Pshirkov:2015tua,Giovannini:2017rbc,Sutton:2017jgr,Jedamzik:2018itu,Paoletti:2018uic} 
and the data from the ultra-high-energy cosmic rays~\cite{Bray:2018ipq} constrain the strength of these fields to be \cite{Fujita:2019pmi,Fujita:2014sna}
\begin{align}
10^{-9}G \gtrsim B_{\rm obs} \gtrsim 10^{-16}G \times
\left\lbrace
\begin{array}{lc}
1   &\lambda_{\rm B} \gtrsim 1 {\rm Mpc}\\
\\
\sqrt{\dfrac{1 {\rm Mpc}}{\lambda_{\rm B}}} &\lambda_{\rm B} \lesssim 1 {\rm Mpc} \, .
\end{array}\right.
\label{B-bound}
\end{align}
The upper and the lower bounds come from the CMB and blazars data, respectively.

On galactic and cluster scales, the observed magnetic fields may be originated from either astrophysical or primordial processes and both scenarios are currently under active considerations  \cite{Giovannini:2017rbc, Subramanian:2015lua, Durrer:2013pga, Giovannini:2003yn,Widrow:2002ud,Grasso:2000wj,Kronberg:1993vk}. A ``seed" magnetic field may be generated by astrophysical mechanisms and then amplified by astrophysical process such as  the galactic dynamo mechanism.  Although this kind of processes can be employed to generate the magnetic fields on galactic and galactic cluster scales, but the generation of magnetic fields with very large correlation length, typically of $\sim$ Mpc scales or larger, is still a mystery in cosmology \cite{wielebinski2005cosmic,beck2012magnetic,Wielebinski:2002xvq,Bernet:2008qp,Bonafede:2010xg,Giovannini:2003yn,Kandus:2010nw,Durrer:2013pga}. 

The above mentioned lower bound and the large correlation lengths may hint towards the primordial origin of 
the cosmological magnetic fields. Indeed, cosmic inflation may be invoked as a working mechanism to generate magnetic fields with large correlation lengths. Inflation is believed  to generate the observed 
large scale structures in universe. The quantum fluctuations associated with the inflaton field are stretched on superhorizon scales which later source the large scale perturbations. With the same mechanism, one may imagine that quantum fluctuations of magnetic fields are stretched beyond the horizon during inflation 
which later seed the observed magnetic fields on cosmos with very large coherent length scales \cite{Turner:1987bw,Ratra:1991bn}. 

Because of the conformal invariance,  the electromagnetic fluctuations can not be enhanced in simple Maxwell theory in an expanding  background \cite{Turner:1987bw,Ratra:1991bn, Parker:1968mv}.  One has to break the conformal invariance in order to prevent the dilution of electromagnetic field during inflation.  A simple way to break conformal invariance is to introduce an  interaction between the electromagnetic field and the  scalar or pseudo scalar inflaton (or a spectator) field or with the curvature scalars \cite{Turner:1987bw,Ratra:1991bn,Garretson:1992vt,Dolgov:1993vg}.  One of the best-studied model of inflationary magnetogenesis  is the so-called Ratra model 
\cite{ Ratra:1991bn, Gasperini:1995dh,Martin:2007ue,Demozzi:2009fu, Kanno:2009ei, Emami:2009vd,  Fujita:2012rb,Bamba:2003av, Barnaby:2012tk,Giovannini:2013rme,Fujita:2013pgp,Ferreira:2013sqa,Ferreira:2014hma,Kobayashi:2014sga, BazrafshanMoghaddam:2017zgx} in which the action contains the non-minimal coupling   $f^2(\phi)F_{\mu\nu}F^{\mu\nu}$ where 
$\phi$ is the inflaton field and $F_{\mu \nu}$ is the electromagnetic field strength.  However, this model  of magnetogenesis suffers from two main problems, the strong coupling problem \cite{Gasperini:1995dh,Demozzi:2009fu,Fujita:2012rb} and  the backreaction problem \cite{Bamba:2003av,Demozzi:2009fu,Kanno:2009ei, Emami:2009vd, Fujita:2012rb}. The strong coupling problem states that the effective coupling constant $f(\phi)^{-1}$ is very large at the early stage of inflation so the perturbative analysis is  not trusted.  The backreaction problem states that the quantum fluctuations of electric field grows rapidly 
during inflation which would spoil inflation too early.  Furthermore, on top of these two problems, one  should also examine  the consistency of the predictions of this setup  with the CMB observation \cite{Barnaby:2012tk,Giovannini:2013rme,Fujita:2013pgp,Ferreira:2013sqa,Ferreira:2014hma, Fujita:2016qab}

In this paper, we revisit the mechanism of primordial magnetogenesis in $f^2F^2$ model taking into account the stochastic effects of electric and magnetic fields perturbations during inflation.  Stochastic formalism is a 
powerful approach to study the quantum fluctuations during inflation \cite{Starobinsky:1986fx,  Sasaki:1987gy, Nambu:1988je, Nakao:1988yi, Rey:1986zk, 
 Mollerach:1990zf, Starobinsky:1994bd, Finelli:2008zg, Finelli:2010sh,  Martin:2011ib, Kawasaki:2012bk, Fujita:2013cna, Fujita:2014tja, Vennin:2015hra,  Nambu:1987ef,   Nambu:1989uf,  Kunze:2006tu, Prokopec:2007ak,   Garbrecht:2013coa,  Burgess:2014eoa, Burgess:2015ajz, Boyanovsky:2015tba,  Boyanovsky:2015jen, Vennin:2016wnk, Assadullahi:2016gkk, Grain:2017dqa, Noorbala:2019kdd}.
Stochastic formalism is an effective theory for the long wavelength parts of the quantum perturbations which are coarse grained on sub-Hubble scales during inflation.  In this approach, the quantum fields are decomposed into the long and short wavelength modes. As the short modes are stretched and leave the Hubble horizon during inflation, they act as classical noises for superhorizon modes with the amplitude $H/2\pi$ in which $H$ is the Hubble expansion rate during inflation.  These quantum kicks can be translated into stochastic forces acting on the classical evolution of the coarse grained fields. 
Therefore, the coarse grained fields are treated as the  classical fields subject to  stochastic noises imposed by  the short modes.

The stochastic formalism has been employed in \cite{Talebian:2019opf}, see also \cite{Fujita:2017lfu},  
to study  the gauge fields perturbations in $f^2F^2$ model of anisotropic inflation. It was pointed out that stochastic effects of gauge fields perturbations can have non-trivial contributions on statistical anisotropies and curvature perturbations. Motivated by these results, one may expect that stochastic effects play important roles in magnetogenesis mechanism in $f^2F^2$ model as well. We show that indeed stochastic effects can  significantly  modify  the previous results for magnetogenesis  in some parameter space of the model. In addition, we provide new insights for the backreaction effects  in the context of stochastic formalism.

The rest of the paper is organized as follows.  In Sec.~\ref{Ratra-Model}, the magnetogenesis mechanism in $f^2F^2$ setup is reviewed. In Sec.~\ref{Sto-auxiliary} the Langevin equations of the electric and magnetic fields are derived  and the  parameters of the evolution of these fields are investigated. In Sec.~\ref{BToday}, by solving the stochastic differential equations discussed in the preceding section, we search the parameter space of the model and calculate  the amplitude of the magnetic field  at present time. In Sec.~\ref{probability}, a probabilistic interpretation for the amplitude of the magnetic fields based on the Fokker-Planck equation is presented. Finally section~\ref{Conclusion} is devoted to the discussion and a summary of our results. The derivations of the correlation functions of the  stochastic noises are presented in appendix A.

\section{The Model}
\label{Ratra-Model}

The model we consider for magnetogenesis is given by the following action 
\begin{eqnarray}
		\mathcal{S} = \int \mathrm{d}^4 x \, \sqrt{-g} \, \bigg[ \dfrac{M_{P}^2}{2} R - \dfrac{1}{2} g^{\mu \nu} \partial_{\mu} \phi  \partial_{\nu} \phi - V(\phi) - \dfrac{f^2(\phi)}{4} F^{\mu \nu} F_{\mu \nu} \bigg] \, ,
	\label{action}
\end{eqnarray}
in which $\phi$ is the inflaton field, $F_{\mu \nu}$ is the electromagnetic field tensor associated with the U(1) gauge field $A^\mu$, $M_P$ is the reduced Planck mass and $R$ is the Ricci scalar. As discussed before, we allow the coupling $f(\phi)$ between the gauge field and the inflaton field. This coupling is added in order to break the conformal invariance such that the electromagnetic fields  survive the exponential expansion during inflation. The specific form of $f(\phi)$ will be given in the following analysis. 

We assume that the electromagnetic fields have no background components. This means that the electromagnetic fields do not contribute to the background energy and they are excited quantum mechanically. The background is given by a spatially flat, Friedmann-Lemaitre-Robertson-Walker (FLRW) universe, described by the line-element
\begin{align}
{\rm d}s^2 = -{\rm d}t^2+a^2(t)~\delta_{ij}~{\rm d}x^i{\rm d}x^j \, ,
\end{align}
where $a(t)$ is the scale factor and $t$ is the cosmic time.

Thanks to the $U(1)$ gauge invariance, we can choose to work in the Coulomb-radiation gauge wherein $A_0 = \partial_i A^i=0$. Varying the action with respect to the matter fields
and neglecting the gravitational backreactions which are sub-leading,  
we obtain the Klein-Gordon and the Maxwell equations,
\begin{align}
&\ddot{\phi} - \dfrac{\nabla^2}{a^2} \phi + 3 H \dot{\phi} + V_{,\phi}(\phi) - \dfrac{f_{,\phi}(\phi)}{f(\phi)} (E^2+B^2) = 0 \,,\label{EoM-Phi} \\
&\dfrac{1}{\sqrt{-g}}\partial_{\mu} \Big(\sqrt{-g}f^2 F^{\mu\nu}\Big)=0 \, .
\label{EoM-A}
\end{align}
Here $H$ represents  the  Hubble  expansion  rate, $H=\dot{a}/ a$, in which a dot denotes the  derivative with respective to cosmic time while the electric and magnetic fields, appearing in Eq.~\eqref{EoM-Phi}, are defined as
\begin{eqnarray}
E_i \equiv -\dfrac{f}{a} \partial_t A_i \,,
~~~~~
B_i \equiv \dfrac{f}{a^2} \epsilon_{ijk} \partial_j A_k \,.
\label{E&B}
\end{eqnarray}

With above definitions, one can obtains the equations of motion for the electric and magnetic fields as
\begin{align}
&\ddot{E}_i - \dfrac{\nabla^2}{a^2} E_i + 5H \dot{E}_i  + \Big[ 6H^2 \Big(1-\dfrac{1}{3}\epsilon_H\Big) + \frac{\ddot{f} + H\dot{f}}{f} - 2\frac{\dot{f}^2}{f^2}\Big]E_i
=0 \,. \label{E_EoM}
\\
&\ddot{B}_i - \dfrac{\nabla^2}{a^2} B_i + 5H \dot{B}_i  + \Big[ 6H^2 \Big(1-\dfrac{1}{3}\epsilon_H\Big) - \frac{\ddot{f} + H\dot{f}}{f} \Big]B_i
=0 \,, \label{B_EoM}
\end{align}
where the slow-roll parameter $\epsilon_H$ is defined as
\begin{align}
\epsilon_H \equiv -\dfrac{\dot{H}}{H^2} \ll 1 \,.
\end{align}
At the background level the expansion rate in the slow-roll limit is  given by 
\begin{align}
3 M_P^2 H^2 \simeq V(\phi) \,,
\label{Friedmann}
\end{align}
 where we have assumed that $V(\phi) \gg \dot{\phi}^2/2$ in order to have a long period of slow-roll inflation.

The conformal coupling is a function of the inflaton field $\phi$ so as the field rolls over the potential, $f$
changes with time. We consider the following phenomenological  ansatz for the conformal coupling
\begin{eqnarray}
f = f_{\rm end}\big(\dfrac{\eta}{\eta_{\rm end}}\big)^{n}\,, ~~~~~~~~~~~~~~ \eta \in (-\infty,0)\, ,
\label{f}
\end{eqnarray}
where $\eta$ represents the conformal time ${\rm d}\eta={\rm d}t/a$, 
$\eta_{\rm end}$ and $f_{\rm end}$ are the values of the conformal time and $f$ at the end of inflation respectively.  The assumption is that the inflaton field decays to radiation at the end of inflation and the conformal factor stabilizes to a fixed value  so we take $f_{\rm end}=1$. The effective gauge coupling is $f^{-1}$ so in order for the perturbative field theory to be applicable we require $n>0$ while the case $n<0$ corresponds to the strong coupling regime.

At the perturbation level the quantum fluctuations of scalar field $\delta \phi$ source the curvature perturbation $\zeta=-H \delta \phi/\dot{\phi}$, generating the following power spectrum for $\zeta$
\begin{align}
{\cal P}_\zeta=\dfrac{H^2}{8\pi^2 M_P^2 \epsilon_H}
\,.
\label{Power-zeta}
\end{align}
Also, the power spectrum of the tensor modes is given by 
\begin{align}
{\cal P}_{\rm t}=\dfrac{2H^2}{\pi^2 M_P^2}
\,.
\label{Power-t}
\end{align}
The ratio of tensor to scaler power spectrum is denoted by $r_{\rm t} \equiv {\cal P}_t/{\cal P}_\zeta$ which is related to the slow-roll parameter via 
\begin{align}
\label{rt}
r_{\rm t} = 16 \epsilon_H \,.
\end{align}
From the CMB observations \cite{Akrami:2018odb,Aghanim:2018eyx} we find that
\begin{align}
{\cal P}_\zeta \simeq 2.1 \times 10^{-9} \,,  \quad \quad 
r_{\rm t} \lesssim 0.056 \, .
\end{align}
Equivalently, these results imply an upper bound on the Hubble parameter during inflation as
\begin{align}\label{H}
H = 2.4 \times 10^{-5}M_P \left( \dfrac{r_{\rm t}}{0.056} \right)^{\frac{1}{2}}\, .
\end{align}
Remember that for the GUT scale inflation we have $H \simeq 10^{-6}M_P$ ($r_{\rm t} \simeq 10^{-4}$). We will occasionally use $r_{\rm t} \simeq 0.01$ and $\epsilon_H \simeq 10^{-3}$ throughout the paper, except mentioned otherwise.

The quantum fluctuations of gauge field $A_i$ are the seeds of large-scale magnetic fields. 
Going to Fourier space, these fluctuations are expanded as 
\begin{eqnarray}
\label{A}
\boldsymbol{A}(\eta,\boldsymbol{x}) &=& \sum_{\lambda = \pm} \int \frac{{\rm d}^3k} {\left(2\pi\right)^3} \,  e^{i\boldsymbol{k}.\boldsymbol{x}} ~\boldsymbol{e}^\lambda(\hat{\boldsymbol{k}}) \left[ A_\lambda(\eta,k) \,\hat{a}^\lambda_{\boldsymbol{k}} +  A_{\lambda}^{*}(\eta,k)\, \hat{a}^{\lambda \dagger}_{-\boldsymbol{k}} \right] \,,
\end{eqnarray}
where  $\boldsymbol{e}^\lambda$ are the circular polarization vectors satisfying the relations
\begin{eqnarray}
\boldsymbol{e}^\lambda(\hat{\boldsymbol{k}}).\boldsymbol{e}^{\lambda'}(\hat{\boldsymbol{k}})&=&\delta^{\lambda\lambda'} \,,\\
\boldsymbol{\hat{k}}.\boldsymbol{e}^\lambda(\hat{\boldsymbol{k}}) &=& 0 \,,\\
i\hat{\boldsymbol{k}} \times \boldsymbol{e}^\lambda &=& \lambda \boldsymbol{e}^\lambda \,,
\label{k-cross-e}
\\
\boldsymbol{e}_\lambda(\hat{\boldsymbol{k}}) &=& \boldsymbol{e}^*_{\lambda}(-\hat{\boldsymbol{k}}) \,,\\
\sum_{\lambda = \pm} e_i^{\lambda}(\hat{\boldsymbol{k}})~e_j^{\lambda}(\hat{\boldsymbol{k}}) &=& \delta_{ij}-\hat{k}_i \hat{k}_j \,.
\end{eqnarray}
Also $\hat{a}^\lambda_{\boldsymbol{k}}$ and $\hat{a}^{\lambda \dagger}_{-\boldsymbol{k}}$ represent the annihilation and creation operators, respectively, satisfying the commutation relation,
\begin{eqnarray}
[\hat a^\lambda_{\boldsymbol{k}}, \hat a^{\lambda' \dagger}_{\boldsymbol{k}'}] &=& (2\pi)^3\delta^{\lambda \lambda'}\delta(\boldsymbol{k} - \boldsymbol{k}') \,.
\end{eqnarray}

Defining the canonically normalized field  $v_\lambda$ as
\begin{align}
\label{Mukhanov-Sasaki}
v_\lambda (\eta,k) \equiv f(\eta) A_\lambda(\eta,k) \,,
\end{align}
the evolution of $v_\lambda$ is given by
\begin{eqnarray}
v_\lambda''+\big( k^2 - \dfrac{f''}{f} \big)v_\lambda =0 \,,
\label{fA-EoM}
\end{eqnarray}
where a prime denotes the derivative with respect to the conformal time $\eta$. Imposing the Bunch-Davies (Minkowski) initial condition for the modes deep inside the horizon, 
\begin{align}
\label{sub-v}
\lim\limits_{\eta \rightarrow -\infty}v_\lambda(\eta,k) \simeq \dfrac{1}{\sqrt{2k}} e^{-i k \eta}\, ,
\end{align}
and using the form of $f(\eta)$ given in Eq. (\ref{f}), the solution is given by 
\begin{eqnarray}
v_{\lambda} = \dfrac{\sqrt{-\pi \eta}}{2} ~H^{(1)}_{n-\frac{1}{2}} (-k \eta) \,,
\label{v-Sol}
\end{eqnarray}
where $H^{(1)}_{\nu} (x)$ is the Hankel function of the first kind.

Inserting Eqs. \eqref{v-Sol} and \eqref{Mukhanov-Sasaki} into Eq. \eqref{E&B}, the electric and magnetic mode functions are given by
\begin{eqnarray}
E_{\lambda} &=&
i\dfrac{\sqrt{\pi}}{2}~k H^2 ~\eta^{5/2} ~H^{(1)}_{n+\frac{1}{2}}(-k\eta) \, ,
\label{E_mode}
\\
B_\lambda &=&
i\dfrac{\sqrt{\pi}}{2}~k H^2 ~\eta^{5/2} H^{(1)}_{n - \frac{1}{2}}(-k\eta) \, .
\label{B_mode}
\end{eqnarray}

The correlation function of the gauge field fluctuations is given by 
\begin{align}\label{A2}
A^2 \equiv
\langle 0| A_i(\eta,\boldsymbol{x})~A^i(\eta,\boldsymbol{x}) |0\rangle = 
\int {\cal P}_A(\eta,k)
~{\rm d}\ln k \, ,
\end{align}
in which  ${\cal P}_A(\eta,k)$ is the dimensionless power spectrum
\begin{align}
{\cal P}_A(\eta,k) \equiv  \dfrac{k^3}{4\pi^2a^2f^2} \sum_{\lambda \equiv \pm} {|v_\lambda(\eta,k)|^2 } \,.
\end{align}
Correspondingly, the power spectra of electric and magnetic fields are given by 
\begin{align}
{\cal P}_E(\eta,k) &=  \dfrac{k^3}{4\pi^2a^4f^2}\sum_{\lambda = \pm} {|v_\lambda'(\eta,k)|^2 }\,,
\label{PowerE}
\\
{\cal P}_B(\eta,k) &=  \dfrac{k^5}{4\pi^2a^4f^2} \sum_{\lambda = \pm} {|v_\lambda(\eta,k)|^2 } \,.
\label{PowerB}
\end{align}

Since  the electromagnetic field is considered as a test fields at the background, the energy density associated with its quantum fluctuations should remain subdominant during inflation. The energy momentum tensor associated with gauge field is given by 
\begin{align}
T^{(A)}_{\mu\nu} = f^2(\phi) \big( F_{\mu\alpha}F_{\nu}{}^\alpha -\dfrac{1}{4}g_{\mu \nu}F_{\alpha\beta}F^{\alpha\beta} \big) \, .
\end{align}
Calculating the expectation value of the energy density, $\langle T^{(A)\, 0}{}_0 \rangle $, we obtain  
\begin{align}
\rho_{\rm em} = \langle 0 |T^{(A)\, 0}{}_0 |0 \rangle &= \dfrac{1}{4\pi^2 a^4} \sum_{\lambda=\pm}\int k^3 ~f^2 \left(|A_\lambda'|^2 + k^2 |A_\lambda|^2  \right)~{\rm d}\ln k
\\
&=
\dfrac{1}{4\pi^2 a^4} \sum_{\lambda=\pm}\int k^3 ~ \left(|v_\lambda'|^2 -\dfrac{f'}{f}|v_\lambda|^2{}'+ \left(k^2+\dfrac{f'{}^2}{f^2}\right) |v_\lambda|^2  \right)~{\rm d}\ln k
\\
&= \dfrac{1}{2}(E^2+B^2) \,,
\end{align}
in which $E^2$ and $B^2$ are defined the same as in Eq.  \eqref{A2}. To control the backreaction of the generated electromagnetic fields on the background, we demand that the ratio of the energy density of electromagnetic fields  $\rho_{\rm em}$ to  inflaton energy  $\rho_\phi \simeq V(\phi)$ remains small  during inflation, 
\begin{align}
R \equiv \dfrac{E^2+B^2}{6 M_P^2 H^2} \ll 1\, ,
\label{R}
\end{align}
where we have used the Friedmann equation \eqref{Friedmann}.


\subsection{Magnetic field in the absence of stochastic effects}
\label{Ratra-today}

Now we estimate the amplitude of magnetic field today which is generated in this setup
in the absence of the stochastic effects. 
 
Suppose the amplitude of magnetic field at the end of inflation with an instant reheating to be 
$B_{\rm end}$. Neglecting the Faraday's  induction in the presence of strong magnetic fields \cite{Kobayashi:2019uqs},  the electromagnetic energy density is diluted like radiation. Then, the strength of the magnetic fields at the present time is given by
\begin{align}
B_{\rm now} =  \left(\dfrac{a_{\rm end}}{a_0}\right)^2 B_{\rm end} \,,
\label{B-now}
\end{align}
where $a_{\rm end}$ and $a_0$ are the values of the scale factor at the end of inflation and at present, respectively. Considering the instant reheating scenario the reheating temperature is given by
\begin{align}
3 M_P^2 H^2 \simeq \dfrac{\pi^2}{30}g_{\rm rel}~T_{\rm rh}^4 \,,
\end{align}
in which $g_{\rm rel}\sim106$ is the relativistic degrees of freedom at the end of reheating. 
Assuming for simplicity that the Universe was radiation dominated throughout  its history, we have 
\begin{align}
\dfrac{a_0}{a_{\rm end}} \simeq \dfrac{T_{\rm rh}}{T_0} \sim 2 \times 10^{28}\left(\dfrac{r_{\rm t}}{0.01}\right)^{1 \over 4}
\,,
\label{a-scale}
\end{align}
where we have used Eq.~\eqref{H} while $T_0 \simeq 2.73 ~{\rm K}$ is the CMB temperature today.   Using these relations, the amplitude of the observed magnetic field at the present time is about
\begin{align}
\label{Bnow2}
B_{\rm now} = 2.5 \times 10^{-57} \left(\dfrac{r_{\rm t}}{0.01}\right)^{-{1 \over 2}}
~B_{\rm end} \, .
\end{align}
Finally, using the definition of the power spectrum of magnetic field ${\cal P}_B $ from Eq. (\ref{PowerB}), the typical amplitude of the mode $k$ of the magnetic field at the end of inflation is given by
\begin{align}
\label{Bend}
B_{\rm end} = \sqrt{{\cal P}_B(\eta_{\rm end},k)} = \dfrac{|v_\pm(\eta_{\rm end},k)| ~k^{5/2}}{\sqrt{2}\pi a^2_{\rm end} f_{\rm end}} \,,
\end{align}
where $v_\pm$ is either of  the two  polarization modes defined in Eq.\eqref{v-Sol} as both polarization 
have equal amplitude. 	

Let us study the behaviour of the mode function $v_\lambda$ at the end of inflation which appears in Eq. (\ref{Bend}).  
Since the cosmological modes of interests are all superhorizon ($k\eta \rightarrow 0$),  the behaviours of these modes can be obtained by using the small argument limit of the Hankel functions,
\begin{align}
\label{Hanckel_limit}
\lim\limits_{x \rightarrow 0}~ H^{(1)}_{\nu}(x) \propto x^{-|\nu|} \,, ~~~~~~~~~\text{for}~~ \nu \neq 0 \,.
\end{align}
Correspondingly, the amplitude of $v_\lambda (\eta,k)$ on superhorizon scales from \eqref{v-Sol} is obtained to be 
\begin{align}
\lim\limits_{k|\eta| \rightarrow 0}~v_\lambda (\eta,k) \simeq C_1~\Theta(\dfrac{1}{2}-n)~ k^{n-{1 \over 2}} \eta^n+ C_2~\Theta(n-\dfrac{1}{2})~k^{-n+{1 \over 2}} ~\eta^{-n+1} \,,
\label{sup-v}
\end{align}
where $\Theta(x)$ is the Heaviside function and 
the constant coefficients $C_i$'s can be obtained by demanding that the mode function \eqref{sub-v} on subhorizon scales connects to the solutions \eqref{sup-v} at the horizon crossing $a_k H = k$.

Having obtained the solution of $v_\lambda (\eta,k)$ on superhorizon scales, we can also calculate the electromagnetic fields on superhorizon scales, obtaining
\ba
\label{E-form}
\lim\limits_{k|\eta| \rightarrow 0}~E_\lambda (\eta,k) &\simeq C_3~\Theta(-n-\dfrac{1}{2})~ k^{n+{3 \over 2}}~\eta^{n+3}+ C_4~\Theta(n+\dfrac{1}{2})~k^{-n+{1 \over 2}} ~\eta^{2-n} \,,
\\
\label{B-form}
\lim\limits_{k|\eta| \rightarrow 0}~B_\lambda (\eta,k) &\simeq C_5~\Theta(\dfrac{1}{2}-n)~ k^{n+{1 \over 2}}~\eta^{n+2}+ C_6~\Theta(n-\dfrac{1}{2})~k^{-n+{3 \over 2}} ~\eta^{3-n} \,,
\ea
in which the coefficients $C_i$ above can be fixed from $C_1$ and $C_2$ in Eq. \eqref{sup-v}.

The first term in Eq. \eqref{sup-v} results in a constant mode for the gauge field $A_\lambda$ while the second term could be a decaying (growing) mode for $n<{1 \over 2}$ ($n>{1 \over 2}$). Demozzi \textit{et al.} \cite{Demozzi:2009fu} have used this result to classify  the $f^2 F^2$ models of primordial magnetogenesis in two categories and studied the related issues. We review their case studies  below. However, before doing that, let us look at specific values of $n$ where the power spectra of electric or magnetic fields become scale invariant, i.e. either $E_\lambda (\eta,k)$ or $B_\lambda (\eta,k)$  from Eqs.  (\ref{E-form}) and (\ref{B-form}) scales like $k^{-3/2}$. A scale invariant electric (magnetic) power spectrum is obtained for $n=2\,  (n=3)$ and $n=-3 \, (n=-2)$, in which the former belongs to the weak coupling regime while the latter is in the strong coupling regime.

\subsubsection{Strong coupling case ($n<{1 \over 2}$)}
\label{Strong}

The strong coupling regime corresponds to  $n<0$ in which the effective gauge coupling $f^{-1}$ is very large at the start of inflation  while approaching  to order of unity at the end of inflation. Therefore, in this regime,  during much of the period of inflation the gauge field sector is strongly interacting and a perturbative analysis in the matter sector (as we treated the electromagnetic field so far) is  not trusted  at all. As studied in \cite{Demozzi:2009fu}, as a subset of strong coupling regime, we consider the case $n<{1 \over 2}$ in which  the first term in Eq. \eqref{sup-v} dominates. In this case the dominant mode is constant, $A_\lambda \propto const$, and the leading contribution to the electromagnetic energy density comes from the magnetic part. By matching the superhorizon solution 
Eq. \eqref{sub-v} with the subhorizon solution  Eq.  \eqref{sup-v} at the horizon crossing $|\eta_k|\simeq k^{-1}$, the constant $C_1$ can be determined, yielding
\begin{align}
\label{v-strong}
v_\lambda \simeq \dfrac{1}{\sqrt{2k}}~\left( \dfrac{a}{a_k} \right)^{-n} 
\,,
\end{align}
where we have taken into account that $\eta \propto a^{-1}$ during inflation and $a_k=k/H$ at the moment of horizon crossing.  

Define $\lambda_{\rm ph} \equiv a_{\rm end}/k$ as the physical wavelength corresponding to the comoving wavenumber $k$ at the end of inflation. Combining  Eqs.  \eqref{v-strong} and \eqref{Bend},   the amplitude of the magnetic fields at the end of inflation is obtained to be 
\begin{align}
\label{Bend-strong}
B_{\rm end} \simeq \dfrac{H^2}{2\pi f_{\rm end}} \left( \dfrac{\lambda_{\rm ph}}{H^{-1}} \right)^{-n-2} \,.
\end{align}
For future reference, we can relate $\lambda_{\rm ph}$ to the (coherence) length scale $\lambda_{\rm B}$ of the magnetic field at the present time via
\begin{align}
\label{lambda}
\lambda_{\rm ph} &=
\dfrac{a_{\rm end}}{k} = \dfrac{a_{\rm end}}{a_{\rm 0}} \lambda_{\rm B} \,. 
\end{align}

Demozzi \textit{et al.} \cite{Demozzi:2009fu} concluded that in the strong coupling regime the magnetic field provides the dominant contribution  to the energy density such that   their  backreactions  may grow too large,  violating the condition $R \ll1$ and terminate inflation quickly.  Specifically, requiring that  inflation lasts at least 75 e-folds they concluded that  one needs\footnote{ We comment that our convention for $n$ differers from that of \cite{Demozzi:2009fu}, by $n \rightarrow -n$. } $n \ge -2.2$ 
in order for the magnetic field energy density does not spoil inflation.

Below we consider various special values of $n$ which were studied in 
\cite{Demozzi:2009fu} and present the estimated values of the magnetic fields on ${\rm Mpc}$ scale in the absence of the stochastic effects. We revisit these cases in Sec.~\ref{BToday} with the effects of stochastic dynamics included. 


\begin{itemize}
	\item $\boldsymbol{n=-2}$
		
As mentioned before, the spectrum of the magnetic field is flat for $n=-2$. Using Eqs. \eqref{Bnow2},  \eqref{Bend-strong} and \eqref{H} with $r_{\rm t} \sim {\cal O}(0.01)$, the strength of the generated magnetic fields today is about
		\begin{align}
		\label{Bflat}
		B_{\rm now} \simeq 1.2 \times 10^{-11}G \,.
		\end{align}
		
	\item $\boldsymbol{n=-2.2}$
	
This is considered  in \cite{Demozzi:2009fu} as the  critical case in which inflation is not destroyed by the magnetic field energy density while for $n < -2.2$ one can not have long enough period of  inflation satisfying the condition $R\ll 1$.  The magnetic fields for $n=-2.2$ have a red spectrum so the largest scale has a dominant contribution to the energy density. Requiring that inflation last at least $75$ e-folds,  the amplitude of the magnetic field today in ${\rm Mpc}$ scale is obtained to be  
	\begin{align}
	\label{B-22-Classic}
	B_{\rm now} \simeq 5.5 \times 10^{-7}G \,.
	\end{align}

\end{itemize}


\subsubsection{Weak coupling case ($n>{1 \over 2}$)}

For this category, the second term in Eq. \eqref{sup-v} dominates and matching Eqs. \eqref{sub-v} and \eqref{sup-v} at the moment of horizon crossing  results in
\begin{align}
\label{v-weak}
v_\lambda \simeq \dfrac{1}{\sqrt{2k}}~\left( \dfrac{a}{a_k} \right)^{n-1} 
\,.
\end{align}
Substituting Eq. \eqref{v-weak} into Eq. \eqref{Bend} we obtain
\begin{align}
\label{Bend-weak}
B_{\rm end} \simeq \dfrac{H^2}{2\pi f_{\rm end}} \left( \dfrac{\lambda_{\rm ph}}{H^{-1}} \right)^{n-3} \,. 
\end{align}
This case corresponds to $A_\lambda \propto a^{2n-1}$ and the main contribution to the electromagnetic energy density comes form the electric field. The effective coupling is growing from a small value at the beginning of  inflation to the order of unity at the end of inflation. Therefore the theory is perturbative throughout inflation.

As shown by Demozzi \textit{et al.} \cite{Demozzi:2009fu}, in the weak coupling regime 
the main contribution to the electromagnetic energy density comes from the electric field 
and their  backreactions may spoil inflation too early. Requiring that inflation lasts at least 75 e-folds, the condition $R<1$ can be satisfied  only if $n \le 2.2$. In the following, we consider various special  cases of $n$ in the weak coupling regime.

\begin{itemize}

\item $\boldsymbol{n=2}$
		
The case $n=2$ correspond to the well known setup of anisotropic inflation which generates anisotropic hair in early Universe  \cite{Watanabe:2009ct,Emami:2013bk,Bartolo:2012sd}, for a review of anisotropic inflation see \cite{Emami:2015qjl} and references therein.   
The electric field fluctuations are nearly constant and nearly scale invariant outside the horizon, while the magnetic field fluctuations rapidly fall off and have a red spectrum,
\begin{eqnarray}
E_\lambda \simeq
{3 H^2 \over \sqrt{2}k^{3/2}} \label{E_superhorizon} \,,
~~~~~~~~
B_\lambda \simeq
{H^2 \eta \over \sqrt{2}k^{1/2}} \label{B_superhorizon}  .
\end{eqnarray}

In the absence of stochastic effects, the strength of magnetic field today in Mpc scale reaches to
		\begin{align}
		\label{B22standardn2}
		B_{\rm now} \simeq 6.3 \times 10^{-35}G \, ,
		\end{align}
		which is too small to work as a seed for a possible dynamo mechanism.

\item $\boldsymbol{n=2.2}$
		
		The magnetic field for $n=2.2$ has a blue spectrum so the small scales have the dominant contributions to the energy density. The amplitude of the magnetic field today in ${\rm Mpc}$ scale is given by
		\begin{align}
		\label{B22standard}
		B_{\rm now} \simeq 2.8 \times 10^{-30}G \,,
		\end{align}
which is again too small  as the seed of primordial magnetic fields. 

\item $\boldsymbol{n=3}$
		
The spectrum of the magnetic field is flat and the strength of the generated magnetic fields today is again given by Eq. \eqref{Bflat}.  However, as mentioned above, the electric field backreaction becomes important terminating inflation quickly.
				
\end{itemize}

The above was a summary of the results for the primordial magnetic field in the setup of (\ref{action}) in the absence  of the stochastic effects. 
In the following  section we revisit this analysis  while taking into account the stochastic effects of the electromagnetic fields which play crucial roles. 
We revisit the conclusion of  \cite{Demozzi:2009fu} that keeping the backreactions under control,  the strength of the primordial magnetic field can not exceed $10^{-30} G$ in Mpc scale today in the weak coupling regime. 


\section{Stochastic dynamics of electromagnetic perturbations}
\label{Sto-auxiliary}

In this section we present the stochastic dynamics of the electromagnetic fields perturbations in details.

Since the forms of electric and magnetic mode functions for both polarization are similar, we combine them  into an auxiliary vector field $X_i$
satisfying the following equation, 
\ba
\ddot{X}_i - \dfrac{\nabla^2}{a^2} X_i + 5H \dot{X}_i {- \left[ (\nu-\dfrac{5}{2})(\nu+\dfrac{5}{2})-2(\nu^2-\dfrac{5}{4})\epsilon_H
+{\cal O}(\epsilon_H^2)
\right]}
H^2 X_i
=0  \label{X_EoM}
\ea
with the following solution for the mode function in Fourier space (assuming the Minkowski initial condition) 
\ba
X(k,\eta) = i\dfrac{\sqrt{\pi}}{2}~k H^2 ~\eta^{5/2} ~H^{(1)}_{\nu}(-k\eta) 
\label{X_mode} \, .
\ea
One can check that Eqs.~\eqref{E_EoM} and \eqref{B_EoM} for the electric and magnetic fields perturbations  are recovered from Eq.~\eqref{X_EoM}  by setting 
{$\nu \rightarrow n+{1 \over 2}$ and $\nu \rightarrow n-{1 \over 2}$}, respectively. Specifically, by taking 
\begin{align}
E_\lambda = X(k,\eta)|_{ {\nu \rightarrow n+{1 \over 2}}} \,,
~~~~~~~~~~~~~~~
B_\lambda = X(k,\eta)|_{{ \nu \rightarrow n-{1 \over 2}} }\,,
\label{EB-X}
\end{align}
the mode function Eq. \eqref{X_mode} reduces to Eqs. \eqref{E_mode} and \eqref{B_mode} for the electric and magnetic fields  respectively. 

Hence, by studying the auxiliary field $X_i$ through Eq.~\eqref{X_EoM}, the evolution of electric and magnetic fields can be found. Our goal is to study the dynamics of $X_i$ using the stochastic formalism. Following the methods of  \cite{Sasaki:1987gy, Nambu:1988je, Nakao:1988yi}, we split the field $\boldsymbol{X}(t,\boldsymbol{x})$ and its conjugate momentum  $\boldsymbol{\Pi}(t,\boldsymbol{x})$ into the long IR and the short  UV modes. The decomposition into the long and short modes  is performed via the Heaviside function $\Theta \left(k-\varepsilon a H\right)$ as a window function. More specifically,  we have  
\ba
\boldsymbol{X}(t,\boldsymbol{x}) &=& \boldsymbol{X}^{\rm IR}(t,\boldsymbol{x}) + \sqrt{\hbar} \int \frac{{\rm d}^3k} {\left(2\pi\right)^3} \, \Theta \left(k-\varepsilon aH\right) \boldsymbol{X}_{\boldsymbol{k}}(t)~ e^{i\boldsymbol{k}.\boldsymbol{x}} \,,\label{X-UV-IR-dec}
\\
\boldsymbol{\dot{X}} (t,\boldsymbol{x}) &=& \boldsymbol{\Pi}_{\boldsymbol{}}^{\rm IR}(t,\boldsymbol{x}) + \sqrt{\hbar}\,\int \frac{{\rm d}^3k} {\left(2\pi\right)^3} \, \Theta \left(k-\varepsilon aH\right) \boldsymbol{\dot{X}}_{\boldsymbol{k}}(t)~ e^{i\boldsymbol{k}.\boldsymbol{x}} \,,\label{XPi-UV-IR-dec}
\ea
in which $\varepsilon$ is a small constant parameter and $\boldsymbol{X}_{\boldsymbol{k}}(t)$ is given by 
\ba
\boldsymbol{X}_{\boldsymbol{k}}(t) &=& \sum_{\lambda = \pm} \boldsymbol{e}^\lambda(\hat{\boldsymbol{k}}) \left[ X_\lambda(t,k)\,\hat{a}^\lambda_{\boldsymbol{k}} +  X_\lambda^{*}(t,k)\, \hat{a}^{\lambda \dagger}_{-\boldsymbol{k}} \right] \,,
\label{X_lambda}
\ea
where the mode function $X_\lambda(t,k)$ is given by \eqref{X_mode}.
Note that the quantum nature of the short modes are indicated by the factor $\sqrt{\hbar}$ in the above expansion.

To investigate the stochastic effects, we expand Eq. \eqref{X_EoM} around $\boldsymbol{X}^{\rm IR}$ and $\boldsymbol{\Pi}^{\rm IR}$ and keep terms up to first order of $\sqrt{\hbar}$. In addition, we discard the term containing the spatial derivatives of the long modes, obtaining   
\begin{align}
\boldsymbol{\dot{\Pi}}^{\rm IR}
&= -5H \boldsymbol{\Pi}^{\rm IR}  {+ \left[ (\nu-\dfrac{5}{2})(\nu+\dfrac{5}{2})-2(\nu^2-\dfrac{5}{4})\epsilon_H
\right]}  H^2\boldsymbol{X}^{\rm IR} +  \sqrt{\hbar}\,\boldsymbol{\tau}
\,,
\label{X_long}\\
\boldsymbol{\dot{X}}^{\rm IR} &= \boldsymbol{\Pi}_{\boldsymbol{}}^{\rm IR} + \sqrt{\hbar}\,~ \boldsymbol{\sigma}  \,, \label{Xdot_long}
\end{align}
in which $(\boldsymbol{\tau}, \boldsymbol{\sigma})$ are the quantum noises, given by
\begin{align}
\boldsymbol{\tau}(t,\boldsymbol{x}) = \varepsilon a H^2 \int \frac{{\rm d}^3k} {\left(2\pi\right)^3} \, \delta \left(k-\varepsilon aH\right) \boldsymbol{\dot{X}}_{\boldsymbol{k}}(t)~ e^{i\boldsymbol{k}.\boldsymbol{x}} \,,\label{tau_Xdot}
\\
\boldsymbol{\sigma}(t,\boldsymbol{x}) = \varepsilon a H^2 \int \frac{{\rm d}^3k} {\left(2\pi\right)^3} \, \delta \left(k-\varepsilon aH\right) \boldsymbol{X}_{\boldsymbol{k}}(t)~ e^{i\boldsymbol{k}.\boldsymbol{x}} \,.\label{sigma_X}
\end{align}

The noises $\boldsymbol{\tau}$ and $\boldsymbol{\sigma}$ arise from the backreactions of the short modes which affect the dynamics of long modes on superhorizon scales. The properties of these noises can be obtained from the behaviour of the auxiliary mode function \eqref{X_mode}, see App. \ref{noise} for more details. The noises $(\boldsymbol{\tau},\boldsymbol{\sigma})$ are stochastic in nature while their quantum non commutativity  disappears on superhorizon scales  by choosing  a sufficiently small parameter $\varepsilon$  so they behave as \textit{classical noises}.

Let us use the number of e-folds, ${\rm d}N=H{\rm d}t$, as the  time variable and define the vectorial normalized white noise $\boldsymbol{\xi}$ as
\ba
\langle \boldsymbol{\xi} (N)\rangle =0 \,, \quad \quad 
\langle \xi_i (N) \xi_j (N')\rangle = \delta_{ij}~ \delta(N-N')  \,.
\ea
Also, let us define the following dimensionless stochastic variable
\begin{eqnarray}
\boldsymbol{\cal X} = \dfrac{\boldsymbol{X}^{IR}}{X_{\rm ref}}
\,, \quad \quad  X_{\rm ref} &\equiv \sqrt{2\epsilon_H}M_P H \, .
\label{calX}
\end{eqnarray}
We show in App. \ref{noise} that the Langevin equation for the long mode can be  cast into a dimensionless stochastic differential equation of the form, 
\begin{eqnarray}
{\rm d}\boldsymbol{\cal X}(N) &=& b_\nu~\boldsymbol{\cal X} ~{\rm d}N+ D_\nu(\varepsilon)~{\rm d}\boldsymbol{W}(N)  \,, \label{calX-Langevin}
\end{eqnarray}
where $\textbf{W}$ is a three dimensional (3D) Wiener processes \cite{evans2013introduction} associated with the noises $\boldsymbol{\xi}$ via
\begin{eqnarray}
\mathrm{d}\boldsymbol{W}(N) &\equiv & \boldsymbol{\xi}(N) ~\mathrm{d}N \,,
\end{eqnarray}
while $b_\nu$ and $D_\nu$ represent the amplitude of the drift and the diffusion terms respectively,  whose specific forms are given in Eqs. (\ref{b}) and (\ref{D}).

Eq.~\eqref{calX-Langevin} is our master equation in the following analysis. Its general solution is given by \cite{evans2013introduction}
\begin{eqnarray}
\boldsymbol{\mathcal{X}}(N) = \boldsymbol{\mathcal{X}}_{\rm cl}(N) +D_\nu(\varepsilon)~e^{b_\nu N}\int_{0}^{N}e^{-b_\nu s} \mathrm{d}\boldsymbol{W}(s)\,,
\label{X-sol}
\end{eqnarray}
where the classical solution $\boldsymbol{\mathcal{X}}_{\rm cl}$, in the absence of stochastic noises, is given by
\begin{eqnarray}
\boldsymbol{\mathcal{X}}_{\rm cl}(N) = \boldsymbol{\mathcal{X}}_{\rm ini} \,e^{b_\nu N} \,,
\label{X_class}
\end{eqnarray}
in which the initial condition $\boldsymbol{\mathcal{X}}_{\rm ini}=\boldsymbol{\mathcal{X}}(0)$ is used. In our setup the assumption is that the electromagnetic fields have no classical background values so $\boldsymbol{\mathcal{X}}_{\rm cl}=0$, but to keep the discussions general we allow for non-zero  initial classical fields values as well.   

Using the following properties of the stochastic integrals \cite{evans2013introduction}
\begin{eqnarray}
\Big\langle  \int_{0}^{T} G(t) {\rm d}W(t)  \Big\rangle = 0 \, , \quad  \quad
\Big\langle \Big[ \int_{0}^{T} G(t) {\rm d}W(t) \Big]^2 \Big\rangle = \Big\langle \int_{0}^{T} G^2 {\rm d}t \Big\rangle \,,
\end{eqnarray}
we can calculate the mean and the variance related to ${\mathcal{X}}_i(N)$. More specifically, 
\begin{eqnarray}
\left\langle 	\mathcal{X}_i(N)  \right\rangle &=& 	 {\mathcal{X}}_{{\rm cl},i}(N)  \,,
\label{X_i}
\\
\left\langle \mathcal{X}_i^2(N)  \right\rangle &=& {\mathcal{X}}_{{\rm cl},i}^2(N)+\dfrac{D_\nu^2(\varepsilon)}{2b_\nu} \left(e^{2b_\nu N}-1 \right) \,,\label{X2}
\\
\delta_{\mathcal{X}_i}^2(N) &=& \dfrac{D_\nu^2(\varepsilon)}{2b_\nu}\left(e^{2b_\nu N}-1 \right) \,,
\\
\langle \mathcal{X}^2(N) \rangle   &=& {\mathcal{X}}_{{\rm cl}}^2(N)+\dfrac{3D_\nu^2(\varepsilon)}{2b_\nu} \left(e^{2b_\nu N}-1 \right)
\,,
\label{X2-aver}
\end{eqnarray}
where the variance is defined via $\delta_\mathcal{X}^2 \equiv \langle  \mathcal{X}^2 \rangle - \langle  \mathcal{X} \rangle^2 $.

Having obtained the magnitude of the auxiliary field in Eq. \eqref{X2-aver} one can investigate the backreactions of the electromagnetic fields on the inflationary background. 
More specifically, the backreaction  generated from the growing electric fields can affect the background energy density and terminate inflation prematurely \cite{Demozzi:2009fu}.  Therefore, we can translate the condition of backreaction 
in terms of the  parameter $R$ defined in Eq. \eqref{R} by requiring  $R \ll1$.   
Using the definition of $X_{\rm ref}$ given in Eq. \eqref{calX}, we can rewrite 
the backreaction condition  as 
\begin{align}
R=\dfrac{1}{3}\epsilon_H \left( \langle {\cal E}^2 \rangle + \langle {\cal B}^2 \rangle \right) \ll 1
\,,
\label{R3}
\end{align}
in which the dimensionless electric and magnetic fields are defined via  
\begin{align}
{\cal E} \equiv \dfrac{E}{X_{\rm ref}} \,,
~~~~~~~~~~~~~~~~~~~~~
{\cal B} \equiv \dfrac{B}{X_{\rm ref}} \,.
\label{EB-Xref}
\end{align}

Alternatively, the Fokker-Planck equation associated  with the Langevin equation \eqref{calX-Langevin} can be employed to describe the time evolution of the probability density function of ${\cal X} (N)$. Consider $f_{{\cal X}_i}(x,N)$ as the probability density function of the random variable ${\cal X}_i$. Then the associated Fokker-Planck equation is given by
\begin{eqnarray}
\dfrac{\partial f_{{\cal X}_i}(x,N)}{\partial N} = -b_\nu\dfrac{\partial}{\partial x} \bigg(x f_{{\cal X}_i}(x,N) \bigg) + \dfrac{D_\nu^2(\varepsilon)}{2} \dfrac{\partial^2}{\partial x^2}f_{{\cal X}_i}(x,N) \,.
\label{probab_X_i}
\end{eqnarray}
Intuitively, one can think of $f_{{\cal X}_i}(x,N) {\rm d}x$ as the probability of ${\cal X}_i$ falling within the infinitesimal interval $[x,x+{\rm d}x]$. 

For a given value of $\nu$ (or $n$), we see that the statistical quantities obtained in Eqs. \eqref{X_i}-\eqref{X2-aver} depend on the number of e-fold $N$, the initial conditions ${\cal X}_{\rm ini}$, the drift $b_\nu$ and the diffusion $D_\nu(\varepsilon)$. Below we investigate how these parameters can  affect  the stochastic properties of the electromagnetic fields.

\subsection{Initial condition }
If the electromagnetic fields have no background classical values then  $\boldsymbol{\mathcal{X}}_{\rm cl}(N)=0$ so  the perturbations ${\cal X}_i$ are generated pure quantum mechanically and 
\begin{eqnarray}
\left\langle 	\mathcal{X}_i(N)  \right\rangle &=& 	 0 \,,
\label{X_i0}
\\
\delta_{\mathcal{X}_i}^2(N)=\left\langle \mathcal{X}_i^2(N)  \right\rangle &=& \dfrac{D_\nu^2(\varepsilon)}{2b_\nu} \left(e^{2b_\nu N}-1 \right) \,,\label{X20}
\\
\langle \mathcal{X}^2(N) \rangle   &=& \dfrac{3D_\nu^2(\varepsilon)}{2b_\nu} \left(e^{2b_\nu N}-1 \right)
\,.
\label{X20-aver}
\end{eqnarray}
Therefore the components of the auxiliary field are described by a pure Brownian motion at early stages,
\ba
\delta_{\mathcal{X}_i}^2 =D_\nu^2(\varepsilon) N \,,~~~~~~~~~~~\big (|b_\nu| N \ll 1 \big) \,.
\ea
The linear growth of the variance with $N$
is the hallmark of the  Brownian motion. We see that even in the absence of a background classical field energy density a large energy density can be generated from stochastic dynamics which can affect the inflationary background as envisaged in \cite{Demozzi:2009fu}. 

 In the rest of the paper we assume $\boldsymbol{\cal X}_{\rm ini}=0$.

\subsection{Drift coefficient }
From equations \eqref{X_i}-\eqref{X2-aver} we see that the sign of $b_\nu$ is very important in determining the fate of the electromagnetic field perturbations  $X_i$. This is the main reason why we kept the $\epsilon_H$ corrections in 
Eq. \eqref{b}, \textit{e.g.} for $\nu=\pm5/2$, we obtain $b_\nu=-\epsilon_H$.

\begin{figure}[t!]
	\begin{subfigure}{0.5\linewidth}
		\centering
		\includegraphics[scale=0.236]{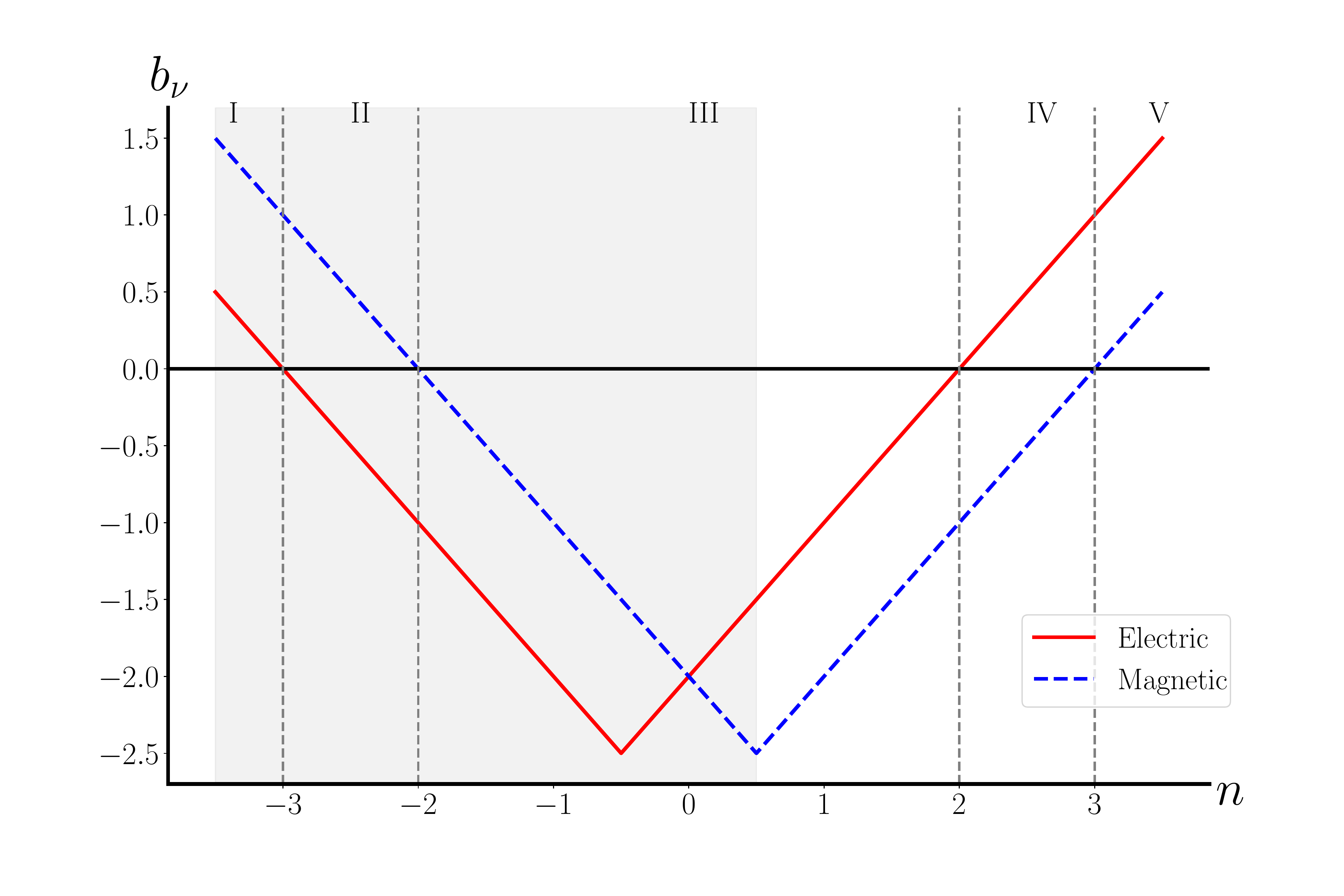}
		\caption{}
		\label{fig:b}
		\vspace{4ex}
	\end{subfigure}
	\begin{subfigure}{0.5\linewidth}
		\centering
		\includegraphics[scale=0.23]{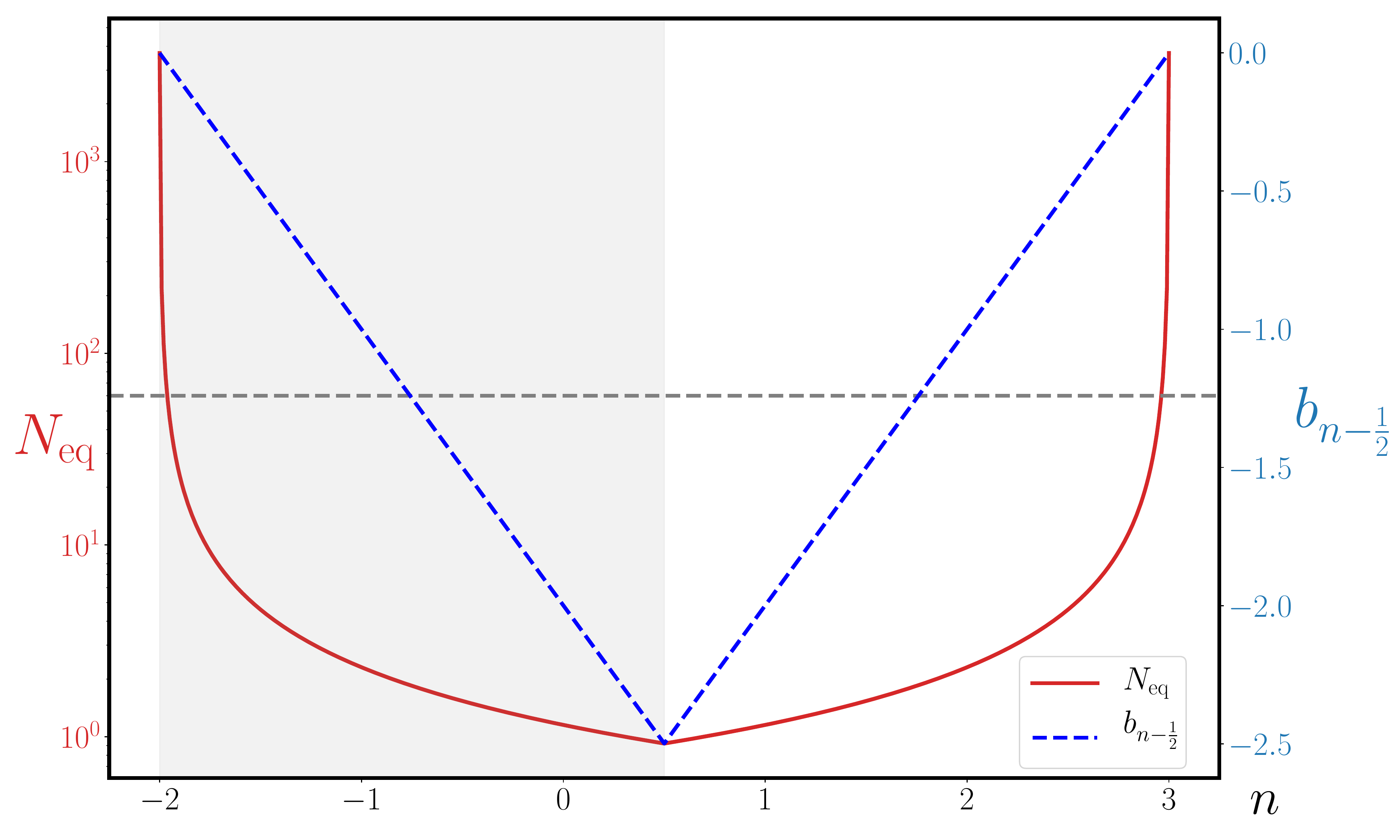}
		\caption{}
		\label{fig:Neqbminus}
	\end{subfigure}
	\caption{(a) $b_\nu$ in terms of $n$ for the electric and magnetic fields, according to Eqs. \eqref{b} and  \eqref{rt} with $r=0.01$. The grey region corresponds to the strong coupling limit while the white regions correspond  to weak coupling limit. We have divided the plot into five regions. In regions  (I) and (V)  we have $b_\nu>0$ for both electric and magnetic fields so the modes grow.  In region (II), corresponding to  $-3<n \leqslant -2$,  the electric field admits an equilibrium state while the magnetic field grows. In region (III) both electric and magnetic field admit stationary state in which the solutions  can be explained by an Ornstein-Uhlenbeck process. In region (IV) electric field grows while magnetic field falls into a stationary state. For each electric and magnetic field there are two points where  $b_\nu=0$ and the 
	solutions are given by a Wiener process. (b) For all ranges of the parameter $n$ shown in this plot $b_\nu<0$ (the right vertical axes) the magnetic field reaches to a stationary regime at the number of e-folds $N=N_{\rm eq}$ given in Eq. (\ref{Neq}).  The grey horizontal dashed line indicates $N_{\rm eq}=60$.}
	\label{fig:Neq}
\end{figure}

As we shall show below, there is a stationary solution for the probability density of ${\cal X}_i$ \eqref{calX-Langevin} if $b_\nu<0$. For $b_\nu=0$ the variance of fluctuations grows linearly with $N$ and the system describes a random walk (Brownian) process while the  exponential growth of the fluctuations takes place for $b_\nu>0$. The behaviour of $b_\nu$ as a function of $n$ is plotted in Fig. \ref{fig:b}.  As an example, consider the region II in this plot, corresponding to $-3<n \leqslant -2$. In this region  the electric  field has $b_{\nu = n+{1 \over 2}}< 0$ so 
it admits a stationary solution for the probability density function  while the magnetic field can grow exponentially ($b_{\nu = n-{1 \over 2}}>0$) or linearly ($b_{\nu = n-{1 \over 2}}=0$). Of course, this range of $n$ falls into the strong coupling regime. 

In the following we have classified the behaviour of the solution of ${\cal X}$  into three categories depending on the sign of $b_\nu$.

\subsubsection{$\mathbf{b_\nu>0}$}

In this case the mean and the variance of the stochastic fields ${\cal X}_i$ grow and there is no stationary probability distribution for its Fokker-Planck equation.
This growing behaviour is linear at early stage when $b_\nu N \ll 1$,
\begin{eqnarray}
\label{X2-averb+}
\langle \mathcal{X}^2 \rangle   \approx
3D^2_\nu(\varepsilon) N \,, \quad \quad 
\delta_{\mathcal{X}_i}^2 \approx& D^2_\nu(\varepsilon) N        \quad \quad (b N <1) \, .
\end{eqnarray}
But when $bN \gtrsim 1$, the stochastic noises grows exponentially and its contributions in total energy density can not be neglected. Let us denote $N= N_{\rm vio}$ as the time  when the energy density of electromagnetic field becomes comparable to the background energy density with large backreactions, violating condition $R \ll 1 $ in Eq.~\eqref{R3}. We then obtain  
\ba
\label{Nvio}
N_{\rm vio} \simeq \dfrac{1}{2 b_\nu} \ln(1+\dfrac{2b_\nu}{ D_\nu^2 \epsilon_H}) \,.
\ea

\subsubsection{$\mathbf{b_\nu=0}$}
\label{sec.b0}

The case $\mathbf{b_\nu=0}$ corresponds to  $|\nu| = \frac{5}{2}+\epsilon_H$, yielding 
\begin{align}
\label{b0}
D_\nu (\varepsilon) &= \sqrt{6 {\cal P}_\zeta} ~ \varepsilon^{-\epsilon_H} \,.
\end{align}
Therefore, the Langevin equation \eqref{calX-Langevin} and the Fokker-Planck equation \eqref{probab_X_i} are  simplified respectively to
\begin{eqnarray}
\label{X_EoM-dimlesbzero}
\mathrm{d}\boldsymbol{\mathcal{X}} &=&  D_\nu(\varepsilon)~\mathrm{d}\boldsymbol{W} \,,
\\
\dfrac{\partial f_{{\cal X}_i}(x;N)}{\partial N} &=&  \dfrac{D^2_\nu(\varepsilon)}{2} \dfrac{\partial^2}{\partial x^2}f_{{\cal X}_i}(x;N) \,.
\label{probab_X_i_b+}
\end{eqnarray}
This situation corresponds to a Wiener process with no drift. The solution of the partial differential equation~\eqref{probab_X_i_b+} is given by
\begin{eqnarray}
\label{f_Xi}
f_{{\cal X}_i}(x;N) = \dfrac{1}{\sqrt{2\pi D^2_\nu(\varepsilon) N}} \mathrm{exp}\Big({-\dfrac{x^2}{2D^2_\nu(\varepsilon) N}}\Big)\,,
\end{eqnarray}
where we have used the initial condition $f_{{\cal X}_i}(x;0)=\delta(x)$. 

The above probability distribution function indicates that ${\cal X}_i$ has a normal (Gaussian) distribution,  denoted by   $\mathbb{N}(0,D^2_\nu(\varepsilon)N)$, describing a random walk process with the variance equal to $D^2_\nu(\varepsilon)N$ and with zero mean.  The probability distribution function for ${\cal X}_i$  obtained in Eq. \eqref{f_Xi}  allows us to extract the  probability density of ${\cal X} = (\sum_i { {\cal X}_i}^2)^{1/2}$ (see App. B in~\cite{Talebian:2019opf} for more details) as
\begin{eqnarray}
\label{f_X_b0}
f_{{\cal X}}(x;N) &=& 2~ \sqrt{\dfrac{1}{2\pi D^6_\nu(\varepsilon)N^3}} ~x^2~  \mathrm{exp} \Big({-\dfrac{x^2}{2D^2_\nu(\varepsilon) N}}\Big)\,.
\end{eqnarray}
Armed with the above probability distribution function, we can compute the associated expectation values and the variance as follows,
\begin{eqnarray}
\left\langle 	\mathcal{X}(N)  \right\rangle &=& \int_{0}^{\infty}{\rm d}x ~x ~f_{{\cal X}}(x;N) 	=D_\nu(\varepsilon) \sqrt{8 N \over \pi} \,,
\label{X_b0}
\\
\langle \mathcal{X}^2(N) \rangle
&=&  \int_{0}^{\infty}{\rm d}x ~x^2 ~f_{{\cal X}}(x;N) = 3D^2_\nu(\varepsilon)N \,,
\label{X-averb0}
\\
\delta_\mathcal{X} ^2(N) &=& (3-\dfrac{8}{\pi})D^2_\nu(\varepsilon)N \,.
\end{eqnarray}

In this Wiener process, we should also check the backreaction effects because the electromagnetic  energy density can grow during inflation and  the condition $R\ll1$ in Eq. \eqref{R3} may be violated. Denoting the time when this condition is violated by $N_{\rm vio}^{\rm W}$, we have 
\begin{align}
\label{NvioWiener}
N_{\rm vio}^{\rm W} \simeq \dfrac{1}{ D_\nu^2 \epsilon_H} \simeq \dfrac{\varepsilon^{2\epsilon_H}}{6{\cal P}_\zeta \epsilon_H}  \,.
\end{align}

The probability distribution functions \eqref{f_X_b0} enables us to calculate the probability of having a given value of  ${\cal X}$ in a desired range. The probability of having ${\cal X}_1<{\cal X}<{\cal X}_2$ at the moment $N$ is given by
\ba
\label{Pb0}
P\left({\cal X}_1<{\cal X}<{\cal X}_2;N\right) &=& \int_{{\cal X}_1}^{{\cal X}_2} {\rm d}x ~ f_{{\cal X}}(x;N)
\\
&=&
\mathrm{Erf}\left(y_2(N)\right)-\mathrm{Erf}\left(y_1(N)\right)
-\dfrac{2}{\sqrt{\pi}} \left(y_2(N)~ e^{-y_2^2(N)}-y_1(N)~ e^{-y_1^2(N)}\right)
\,.
\nonumber
\ea
Here $\mathrm{Erf}$ is the error function and $y_i(N)\equiv \dfrac{{\cal X}_i}{\sqrt{2 N}D_\nu}$ where $i=1,2$. The dependency of the probability density to $N$ is due to the Wiener process describing  a random walk in which the  variance grows linearly with $N$.
\subsubsection{$\mathbf{b_\nu<0} $}
\label{bmin}
The conditions
\begin{align}
b_\nu <0 \,,~~~ D_\nu>0 \,,
\label{bD}
\end{align}
converts Eq.~\eqref{calX-Langevin} into an Ornstein-Uhlenbeck (OU) stochastic differential
equation~\cite{evans2013introduction},
\begin{eqnarray}
\dfrac{{\rm d}\boldsymbol{\cal X}(N)}{{\rm d}N} &=& -|b_\nu|~\boldsymbol{\cal X}+ D_\nu(\varepsilon)~\boldsymbol{\xi}  \,. 
\label{calX-Langevin2}
\end{eqnarray}
Not only  the frictional drift force $-|b_\nu| 	\boldsymbol{\mathcal{X}}$ can balance the random force $D_\nu \boldsymbol{\xi}$~, but also it washes out the explicit dependence of the mean to the initial conditions $\boldsymbol{\mathcal{X}}_{\rm ini}$ over time. It means that the distribution of $\mathcal{X}_i$ approaches the normal distribution $\mathbb{N}\Big(0,\dfrac{D^2_\nu}{2|b_\nu|}\Big)$ as $N \rightarrow \infty$.

The OU process~\eqref{calX-Langevin2} is a stationary Gauss-Markov process in which there is the tendency for the system for drifting toward the mean value, with a greater attraction when the process is further away from the mean. Therefore the field $\boldsymbol{\mathcal{X}}$ admits a stationary probability distribution, $\partial f^{\rm eq}_{{\cal X}_i}/\partial N=0$, with a long-term mean  and a bounded variance (mean-reverting process). The stationary solution of Fokker-Planck Eq. \eqref{probab_X_i} is given by
\begin{eqnarray}
\label{f_Xi_station}
f^{\rm eq}_{{\cal X}_i}(x) = \sqrt{\dfrac{|b_\nu|}{\pi D^2_\nu }} ~\mathrm{exp} \Big({-\dfrac{|b_\nu|}{D^2_\nu}}x^2 \Big)\,.
\end{eqnarray}
Using the above probability density function for the components ${\cal X}_i$, it is easy to obtain the density  function of its magnitude ${\cal X}$
as follows:
\begin{eqnarray}
\label{f_X_station}
f^{\rm eq}_{{\cal X}}(x) &=& 4~ \sqrt{\dfrac{|b_\nu|^3}{\pi D^6_\nu}} ~x^2~   \mathrm{exp} \Big({-\dfrac{|b_\nu|}{D^2_\nu}}x^2\Big)\, .
\end{eqnarray}
This density function allows us to calculate various expectation values and variance associated with ${\cal X}$ as follows:
\begin{eqnarray}
\left\langle \mathcal{X}  \right\rangle_{\rm eq} &=& \int_{0}^{\infty} {\rm d}x~x~f^{\rm eq}_{{\cal X}}(x)=\dfrac{2D_\nu}{\sqrt{\pi |b_\nu|}}\,,
\label{X_bmin}
\\
\langle \mathcal{X}^2 \rangle_{\rm eq} &=& \int_{0}^{\infty} {\rm d}x~x^2~f^{\rm eq}_{{\cal X}}(x)=\dfrac{3D^2_\nu}{2|b_\nu|} \,,
\label{X-averbmin}
\\
\delta_{\mathcal{X}_{\rm eq}} ^2 &=& (\dfrac{3}{2}-\dfrac{4}{\pi})\dfrac{D^2_\nu}{|b_\nu|} \,.
\label{X-varbmin}
\end{eqnarray}
Moreover, we can estimate the equilibrium time when the  field reaches to its stationary value using $f^{\rm eq}_{{\cal X}}(x)$. Let us define  $N_{\rm eq}$ as the time when
$\langle \mathcal{X}^2(N_{\rm eq}) \rangle \rightarrow \langle \mathcal{X}^2 \rangle_{\rm eq}$. Formally, $N_{\rm eq} \rightarrow \infty$, but for practical purposes we can consider $N_{\rm eq}$ as the time when the ratio $\left| \langle \mathcal{X}^2(N_{\rm eq}) \rangle -\langle \mathcal{X}^2 \rangle_{\rm eq} \right|/\langle \mathcal{X}^2 \rangle_{\rm eq}$ drops to a small value say $10^{-2}$.
With this approximation, and using  Eqs.~\eqref{X20-aver} and \eqref{X-averbmin},
we obtain
\begin{align}
	N_{\rm eq} = \dfrac{\ln 10}{|b_\nu|} \, ,
	\label{Neq}
\end{align}
which is plotted for $\nu=n-1/2$ (magnetic field) in Fig.~\ref{fig:Neqbminus}.

It is interesting to check when the backreactions become important, violating the condition 
$R \ll 1$.  
Denoting $N_{\rm vio}^{\rm -}$ as the time when this condition is violated, we obtain
\ba
\label{Nvio0}
N_{\rm vio}^- \simeq \dfrac{-1}{2 |b_\nu|} \ln(1-\dfrac{2|b_\nu|}{ D_\nu^2 \epsilon_H}) \,.
\ea

Using Eq. \eqref{f_X_station}, the probability of the field ${\cal X}$ acquiring a value in the interval ${\cal X}_1<{\cal X}<{\cal X}_2$ is given by
\begin{align}
\label{Pbminus}
P_{\rm eq}\left({\cal X}_1<{\cal X}<{\cal X}_2\right) &= \int_{{\cal X}_1}^{{\cal X}_2} {\rm d}x ~ f^{\rm eq}_{\cal X}(x)
\nonumber
\\&=
\mathrm{Erf}\left(y_2\right)-\mathrm{Erf}\left(y_1\right) -\dfrac{2}{\sqrt{\pi}} \left(y_2~ e^{-y_2^2}-y_1~ e^{-y_1^2}\right)
\,,
\end{align}
in which $y_i\equiv \dfrac{\sqrt{|b_\nu|}}{D_\nu}{\cal X}_i$ and $i=1,2$~.  
\subsection{Diffusion coefficient 
}
\label{DiffSec}
Now we study the effects of the diffusion coefficient $D_\nu(\varepsilon)$.
As mentioned before, the quantum non-commutativity of $(\boldsymbol{\tau},\boldsymbol{\sigma})$ is proportional to $\varepsilon^5$ (see App. \ref{noise} for more details) so they become classical noises when $\varepsilon$ is chosen small enough. Therefore,  it is always possible to consider $(\boldsymbol{\tau},\boldsymbol{\sigma})$ as classical noises while for a fixed $\nu$ the diffusion coefficient is only a function of the small parameter $\varepsilon$. Therefore the dependency of the solutions to the diffusion coefficient is determined by the  value of $\varepsilon$ as given in Eq. (\ref{Dve}).

In single field inflationary scenarios, the coarse graining is over subhorizon modes with  $k \gtrsim \varepsilon a H$ in which  $\varepsilon$
is a cutoff parameter satisfying $e^{-1/3\epsilon_H}\ll \varepsilon \ll 1$ \cite{Sasaki:1987gy, Nambu:1988je, Nakao:1988yi} . Under this condition, in single field inflationary models 
the physical results are independent of $\varepsilon$.
Here, however, the results in general depend on the value of $\varepsilon$.

There are two points here that we elaborate in details. First, the power of $\varepsilon$ in $D_\nu(\varepsilon)$ is directly related  to the scale dependency of the power spectrum of the fields. To see this specifically, let us first consider the magnetic field, corresponding to $\nu = n-1/2$. In the conventional treatment with no stochastic effects considered,  the power spectrum of magnetic  field for $n>1/2$ from Eq. (\ref{Bend-weak})
is $\sqrt{{\cal P}_B} \propto k^{3-n}$ while for $n<1/2$, from Eq. (\ref{Bend-strong}), it is $\sqrt{{\cal P}_B} \propto k^{n+2}$. Now one can check that the power of 
$k$ in either cases are the same as the power of $\varepsilon$ in $D_\nu(\varepsilon)$. The same conclusion applies to electric field as well.  As specific examples,  we have seen that for $n=3$ the magnetic field is scale invariant ($\sqrt{{\cal P}_B} \propto k^0$), while the electric field has a red power ($\sqrt{{\cal P}_E} \propto k^{-1}$). On the other hand, in the stochastic approach, the diffusion coefficients of the magnetic and electric fields are given by  $D_{5/2} \propto \varepsilon^{0}$ and $D_{7/2} \propto \varepsilon^{-1}$ respectively as expected. 
As a result,  from Eq.~\eqref{Dve} we conclude that the electromagnetic  fields have a red spectrum for $|\nu| > 5/2$, a blue spectrum for $|\nu| < 5/2$ and a flat spectrum if $|\nu| = 5/2$. In addition, comparing the power of $\varepsilon$ in $D_\nu(\varepsilon)$ with the 
  sign of $b_\nu$ in Eq. (\ref{b})
  we arrive at another important 
conclusion: the sign of the power of $\varepsilon$ in $D_\nu(\varepsilon)$ is opposite to the sign of $b_\nu$; {\it electromagnetic perturbations  with a blue spectrum have 
 $b_\nu <0$  so they always fall into a stationary state}.

\begin{figure}[t!]
	\includegraphics[width=\linewidth]{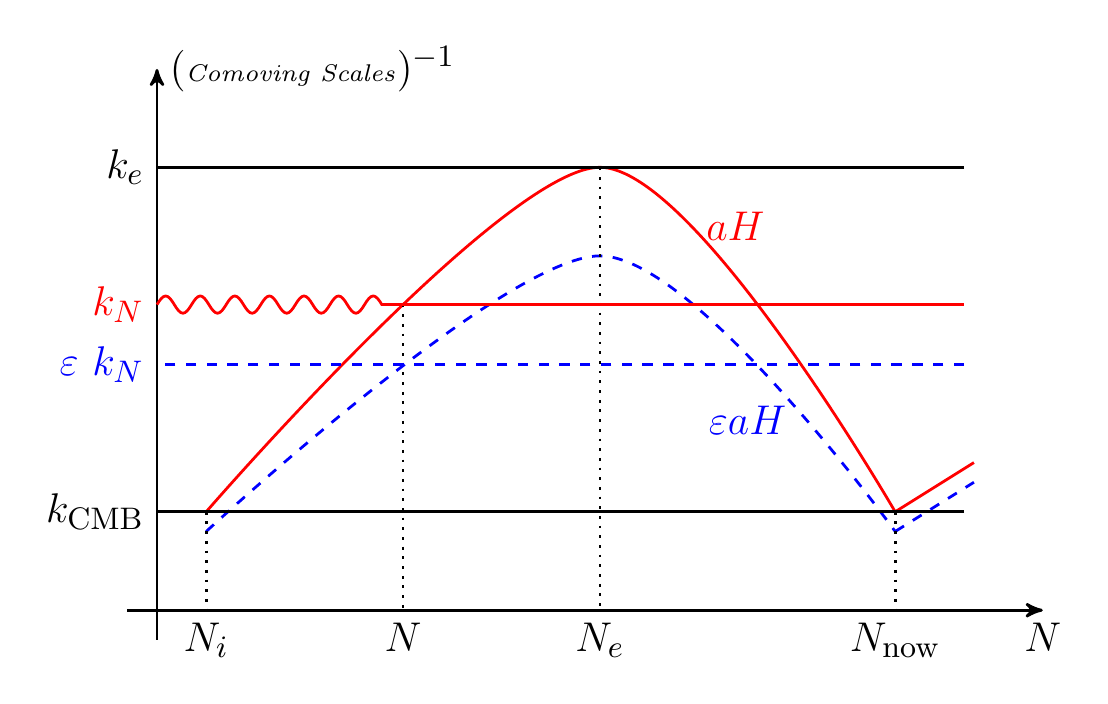}
	\caption{Interpretation of long and short modes in terms of $\varepsilon$. $N_i$ represents the start of the (observed) inflation in which the longest observable mode ($k_{\mathrm{CMB}}$) have left the horizon. $N_e$ is the time of end of inflation and $N_{\mathrm{now}}$ is the current time.  If $\varepsilon=1$, then the coarse graining for a mode of interest $k_N$ will be over the modes in the interval $k_N<k< k_e$, i.e. for the modes inside the Hubble patch. But in the stochastic formalism, the coarse graining is over a wider range, for the modes in the interval $\varepsilon k_N<k< k_e$. }
	\label{fig:varep}
\end{figure}

Second, Eq.~\eqref{Dve} indicates that by choosing small enough value of $\varepsilon$ 
the diffusion coefficient can be very large for $|\nu|>5/2$.  Hence a relevant question is  how small $\varepsilon$ can be? 
The parameter $\varepsilon$ appears in the  window function in Eq.~\eqref{X-UV-IR-dec} when performing the long and short decomposition. When constructing a coarse grained field for the long mode perturbations, depending  on the value of $\varepsilon$, 
all modes smaller than Hubble patch and a fraction of superhorizon modes are integrated out by using the  window function. Specifically, all modes in the range  $k > \varepsilon a H$ are integrated out. If one choose the simplest choice 
$\varepsilon=1$, then the coarse graining will be only on a Hubble patch, i.e. for subhorizon modes. But for  $\varepsilon<1$, not only the subhorizon modes but also a fraction of superhorizon modes  which fall into the window function are also integrated out. The smaller is $\varepsilon$, the larger is this fraction. 
A lower bound (or cut off value)  for $\varepsilon$ can be obtained by integrating out the  largest wavelength observable in CMB, $k_{\rm CMB}$,  in the coarse graining  process. Let us define  $k_N$ as the mode $k$
which leaves the horizon at the number of e-fold $N$. Then  the smallest value for $\varepsilon_{\rm }$ associated to the mode mode $k_N$ 
is given by 
\begin{align}
\varepsilon_{\rm } \equiv \dfrac{k_{\rm CMB}}{k_N}
\end{align}
which is illustrated in Fig.~\ref{fig:varep}.

We may use the Planck observation's pivot scale $k_* = 0.05~{\rm Mpc}^{-1}$ \cite{Akrami:2018odb} as a representative value of $k_{\rm CMB}$. Then for the physical length scale ${\rm 1\,  Mpc}$ today as the coherent length of primordial magnetic fields ($\lambda_{\rm B}=1 {\rm Mpc}$), 
the value of $\varepsilon$ is obtained to be 
\begin{align}
\label{ec}
\varepsilon_{\rm Mpc} = \dfrac{0.05~ {\rm Mpc}^{-1}}{2\pi/1 {\rm Mpc}} = 8 \times 10^{-3} \, .
\end{align}
However,  as mentioned above, a smaller value for $\varepsilon$ can be calculated by considering the largest observable scales today, $k_{\rm CMB} \simeq 10^{-4} {\rm Mpc}^{-1}$ \cite{Akrami:2018odb} which results to $\varepsilon_{\rm Mpc}=1.6 \times 10^{-5}$. Therefore we consider $\varepsilon_{\rm Mpc}$ in the range $10^{-2}-10^{-5}$ in our analysis in next section. For convenience, we do not write the subscript ${\rm Mpc}$
in $\varepsilon_{\rm Mpc}$ and simply use $\varepsilon$ in our analysis.

\section{Magnetic Fields Today}
\label{BToday}

In this section, we calculate the present value of the magnetic fields produced in this model taking into account the stochastic dynamics.  We will see that the amplitude of the generated magnetic field depends on parameter $n$ appearing in the definition of coupling function $f$, Eq.~\eqref{f}. As mentioned before, there are two regimes for this parameter; $n>1/2$ (\textit{weak coupling}) in which one faces with the backreaction problem and $n<1/2$ (\textit{strong coupling}) where the perturbative approaches break down  at the early stage of inflation and  the analysis can not be trusted. 


The present value of the magnetic field $B_{\rm now}$ is related to the magnetic field at the end of inflation via Eq. \eqref{B-now} in which $B_{\rm end}$ itself is determined  from the stochastic  quantity ${\cal B}_{\rm end}$ via the definition \eqref{EB-Xref}, as $B_{\rm end}={\cal B}_{\rm end}~ X_{\rm ref}$. 
The relations derived in the preceding section enable us to calculate
\begin{align}
\label{BendSto}
{\cal B}_{\rm end} = \sqrt{\langle {\cal B}^2(N_{\rm end}) \rangle} \,,
\end{align}
for different cases of drift coefficients $b_\nu>0$, $b_\nu=0$ and $b_\nu<0$ using  Eqs. \eqref{X20-aver}, \eqref{X-averb0} and \eqref{X-averbmin}  respectively. On the other hand,  using the definition of $X_{\rm ref}$ , we obtain
\begin{align}
X_{\rm ref} &\simeq 1.12 \times 10^{50} ~G ~ \left(\dfrac{r_{\rm t}}{0.01}\right)
\,,
\label{X_ref}
\end{align}
where Eqs.~\eqref{rt} and \eqref{H} have been used. Then the present value for the magnetic fields, \eqref{B-now}, is given by
\begin{align}
B_{\rm now} \simeq 2.6 \times 10^{-7} G ~\left(\dfrac{r_{\rm t}}{0.01}\right)^{1 \over 2}~{\cal B}_{\rm end} \,.
\label{Bnow}
\end{align}
As seen, for $r_{\rm t}={\cal O}(0.01)$, the observational bound given by \eqref{B-bound} is roughly satisfied if ${\cal B}_{\rm end}$ satisfies the following constraint
\begin{align}
3.8 \times 10^{-3} \gtrsim {\cal B}_{\rm end} \gtrsim 3.8 \times 10^{-10} \times
\left\lbrace
\begin{array}{lc}
1   &\lambda_{\rm B} \gtrsim 1 {\rm Mpc}\\
\\
\sqrt{\dfrac{1 {\rm Mpc}}{\lambda_{\rm B}}} &\lambda_{\rm B} \lesssim 1 {\rm Mpc}
\end{array}\right.
\,
\label{B-end-bound}
\end{align}

In the following, we  use the estimation $r_{\rm t} \simeq 0.01$ for our calculations and simulations.  The results of this section is summarized in subsection \ref{sum} and in Fig.~\ref{fig:RBnow}.

\subsection{Strong coupling regime}

Although in the regime  $n < 1/2$  the backreaction effects of the electric fields are under control but for $n < 0$ the gauge coupling $f^{-1}$ is incredibly large at the beginning of inflation and the perturbative analysis are not trusted. However, since the dominant contribution to the energy density comes from the magnetic field, it is interesting to investigate the stochastic effects in the spacial cases $n =-2,-2.2$ as well as $n =-2 - \epsilon_H,-3 - \epsilon_H$ in some more details. The last two cases represent Wiener process for the magnetic and electric fields respectively.

\subsubsection{$n=-2$}
\label{n=-2}
For $n=-2$, both electric and magnetic fields have $b_\nu <0$ so their evolutions   are described via the OU processes. These behaviours can be seen from the stochastic differential equations of magnetic ($\nu=-5/2$) and electric ($\nu=-3/2$) fields which are given by
\begin{eqnarray}
{\rm d}\boldsymbol{\cal B} &=& -\epsilon_H~ \boldsymbol{\cal B}~{\rm d}N + \sqrt{6{\cal P}_\zeta}~{\rm d}\boldsymbol{W}  \,, \label{B-2}
\\
{\rm d}\boldsymbol{\cal E} &=& -\boldsymbol{\cal E} ~{\rm d}N+\dfrac{5\sqrt{{\cal P}_\zeta}}{2\sqrt{6}}~\varepsilon~{\rm d}\boldsymbol{W}   \,. \label{E-2}
\end{eqnarray}
In this case, the electric and the magnetic fields have a blue and a scale invariant spectra, respectively. Since  the sign of the drift coefficients are negative 
 both  electric and magnetic fields admit  stationary regimes with the terminal values
\begin{eqnarray}
\langle \mathcal{B}^2 \rangle_{\rm eq} &\simeq& 3 \times 10^{-5} \,,
\label{B2n-2}
\\
\langle \mathcal{E}^2 \rangle_{\rm eq} &\simeq& 2.2 \times 10^{-13} \,,
\end{eqnarray}
where Eq. \eqref{X-averbmin} has been used. Using Eq. \eqref{Neq}, these values of
the electric and magnetic fields are reached 
at around $N_{\rm eq} \simeq 3700$ and $N_{\rm eq} \simeq 3$ e-folds respectively. Assuming that inflation lasts about 3700 e-folds number, and using Eq. \eqref{Bnow},  the present value of the scale invariant magnetic field is given by
\begin{align}
B_{\rm now} \simeq 1.5 \times 10^{-9}~G 
\,.
\label{Bnown-2}
\end{align}
As seen from the above result, this value is larger by 2 orders of magnitude compared to \eqref{Bflat} estimated in conventional approach where the stochastic effects are neglected.

\begin{figure}[t!]
	\begin{subfigure}{0.5\linewidth}
		\centering
		\includegraphics[scale=0.23]{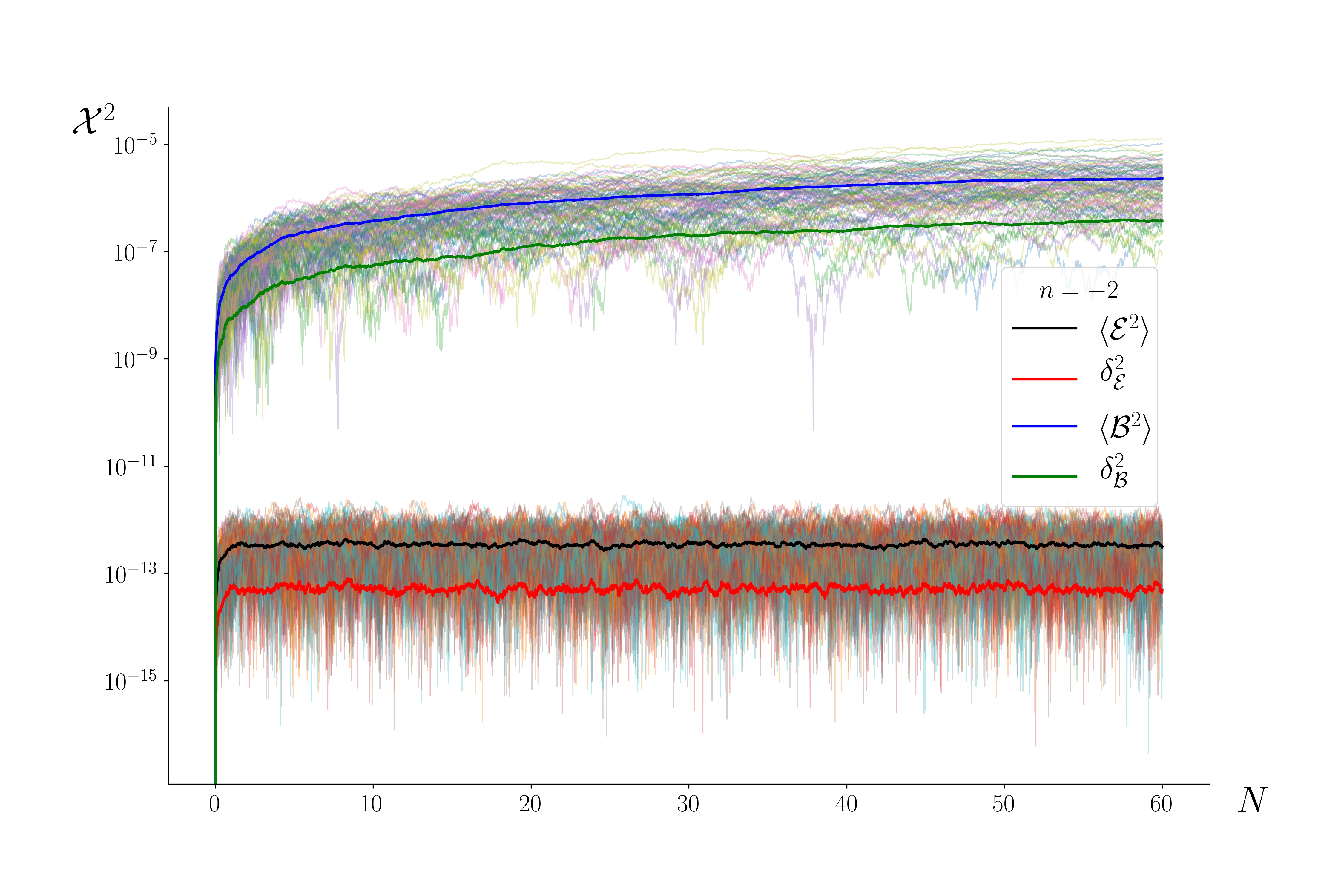}
		\caption{}
		\label{fig:Xn-2}
	\end{subfigure}
	\begin{subfigure}{0.5\linewidth}
		\centering
		\includegraphics[scale=0.23]{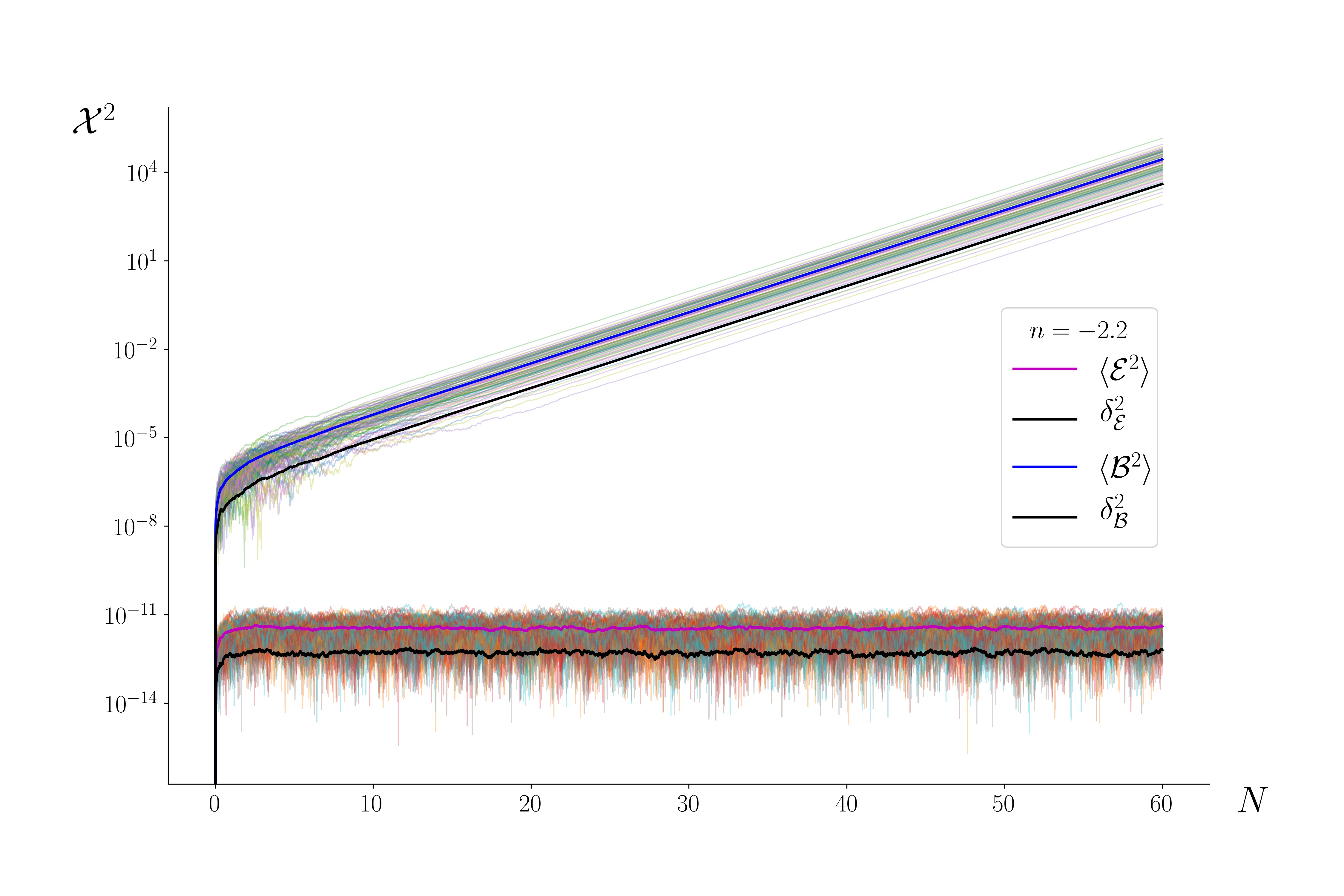}
		\caption{}
		\label{fig:Xn-22}
	\end{subfigure}
	\caption{Simulations of the evolution of $\langle {\cal X}^2 \rangle$ in terms of e-folding  number $N$ in strong coupling regime for one hundred realization with $\varepsilon=10^{-3}$ for~(a) $n=-2$ and (b) $n=-2.2$~.}
	\label{fig:n2}
\end{figure}

Although the backreaction condition~\eqref{R3} is satisfied ($R \simeq 6.6 \times 10^{-9}$), but as  mentioned before, the strong coupling regime raises serious concern about the applicability of the perturbative results. The behaviours of the electric and magnetic fields are shown in Fig.~\ref{fig:Xn-2}.

\subsubsection{$n=-2.2$}
\label{n-2.2}
In this case, the stochastic differential equations for the magnetic field ($\nu=-2.7$) and electric field ($\nu=-1.7$)  are given by
\begin{eqnarray}
{\rm d}\boldsymbol{\cal B} &=& 0.2~ \boldsymbol{\cal B}~{\rm d}N  + 3.4\sqrt{{\cal P}_\zeta}~\varepsilon^{-1/5}~{\rm d}\boldsymbol{W}  \,, \label{B-2.2}
\\
{\rm d}\boldsymbol{\cal E} &=& -0.8~ \boldsymbol{\cal E} ~{\rm d}N  + 1.14 \sqrt{{\cal P}_\zeta}~\varepsilon^{4/5}~{\rm d}\boldsymbol{W} \,. \label{E-2.2}
\end{eqnarray}
The power spectra of the electric and  magnetic fields are blue and red tilted respectively. The electric field admits a stationary value $\langle \mathcal{E}^2 \rangle_{\rm eq} \simeq 2.3 \times 10^{-12}$ at around $N_{\rm vio} \simeq 3$ e-fold while the magnetic field grows exponentially and spoils the condition \eqref{R3} at around $N_{\rm vio}\simeq 55$ e-fold.

Requiring inflation lasts at least 60 e-folds, and using Eq. \eqref{ec} for the value of 
$\varepsilon$,    we obtain
\begin{align}
\langle {\cal B}^2(60)\rangle \simeq 3 \times 10^{4} \,,  \quad \quad 
R \simeq 6.6 \,.
\end{align}
This should be compared with  the results obtained in subsection \ref{Strong}  in the conventional approach in the absence of the stochastic effects where it is concluded that  backreactions are not important. 
Here, however,  in the presence of stochastic effects the backreactions of the magnetic fields  spoil inflation. Also the present value of the magnetic field on ${\rm Mpc}$ scale is about 2 orders of magnitude larger than the value obtained in \eqref{B-22-Classic}. More specifically, using the value of $\varepsilon$ in 
Eq. \eqref{ec},  we obtain 
\begin{align}
B_{\rm now} &\simeq
4.6 \times 10^{-5}~G \, .
\end{align}
The behaviours of the electric and magnetic fields are shown in Fig.~\ref{fig:Xn-22}.

\subsubsection{Wiener processes in Strong coupling regime}

According to subsection \ref{sec.b0}, we  have  a Wiener process for the magnetic field if $n=-2-\epsilon_H$ which leads to
\begin{align}
{\rm d}\boldsymbol{\cal B} &= \sqrt{6 {\cal P}_\zeta} \varepsilon^{-\epsilon_H}~{\rm d}\boldsymbol{W}  \,. \label{n-2b0}
\\
{\rm d}\boldsymbol{\cal E} &= -\boldsymbol{\cal E}~ {\rm d}N+\dfrac{5\sqrt{{\cal P}_\zeta}}{2\sqrt{6}}~\varepsilon~{\rm d}\boldsymbol{W}   \,.
\end{align}
This means that the electric field reaches to the terminal value $\langle {\cal E}^2 \rangle \simeq 2.2 \times 10^{-13}$ at around $N_{\rm eq} \simeq 3$ e-fold while the strength of magnetic field at the end of inflation  is given by
\begin{align}
\label{BWiener-end}
{\cal B}_{\rm end} = \sqrt{\langle \mathcal{B}^2(N_{\rm end}) \rangle} \simeq 2 \times 10^{-4} \sqrt{N_{\rm end}}~ \varepsilon^{-\epsilon_H} \, ,
\end{align}
where  \eqref{X-averb0} has been used. From Eq. \eqref{NvioWiener},  this value of magnetic field can spoil inflation after $N_{\rm vio}^{\rm W} \simeq 1.2 \times 10^{11}$ e-folds. The ratio of the electromagnetic to inflaton energy density is about $R\simeq4\times10^{-10}$ if we assume inflation  lasts $60$ e-folds.

The present value  of the magnetic fields on ${\rm Mpc}$ scale from Eq.~\eqref{Bnow}  is obtained to be  
\begin{align}
\label{BWiener-now}
B_{\rm now} \simeq 5.2 \times 10^{-11}~\sqrt{N_{\rm end}}~G
\,.
\end{align}
This shows  that the amplitude of the magnetic fields today depends on the number of e-folds that inflation was in progress. By choosing $N_{\rm end}=60$, we obtain $B_{\rm now} \simeq 4 \times 10^{-10}~G$.

Also, a Wiener process occurs for the electric field if we choose $n=-3-\epsilon_H$,
\begin{align}
{\rm d}\boldsymbol{\cal B} &= \boldsymbol{\cal B} ~{\rm d}N+\dfrac{35\sqrt{{\cal P}_\zeta}}{\sqrt{6}~\varepsilon}~{\rm d}\boldsymbol{W}   \,, \label{n-3b0B}
\\
{\rm d}\boldsymbol{\cal E} &= \sqrt{6 {\cal P}_\zeta} \varepsilon^{-\epsilon_H}~{\rm d}\boldsymbol{W}  \,. \label{n-3b0E}
\end{align}
In this case a very  large magnetic field is generated at the early stage of inflation,  spoiling inflation at around $N_{\rm vio}^W \simeq 6$ e-folds.

We emphasis that the above results can not be trusted due to strong coupling problem.  But, nonetheless, the  models have interesting aspects from the stochastic points of view where we also compared our results with those obtained via the conventional approach  where the stochastic effects  are neglected.

\subsection{Weak coupling regime}

In this regime, the gauge coupling $f^{-1}$  is exponentially small at the beginning of inflation and grows to the order of unity at the end of inflation so the perturbative analysis is trusted during entire period of inflation. Following the logic of  \cite{Demozzi:2009fu}  we define this regime as $n>1/2$ although to have a weak coupling  one actually requires $n>0$. In this regime, the main contribution of electromagnetic energy density comes from the electric part. Therefore, the backreaction effects could destroy inflation, violating the condition $R \ll 1$  at the
e-folding number given in Eq. \eqref{Nvio} for a wide range of parameter $n$. However, as shown by \cite{Demozzi:2009fu}, there are some range of parameters 
where the backreaction is under control though  the generated magnetic fields are very small, see for example Eq. \eqref{B22standard}. Here we revisit this issue in the presence of the stochastic noises.

In the following, we consider the spacial cases $n=2, n=2.2, n= 3$ and $n=3+\epsilon_H \,, n=2+\epsilon_H$ with more details and simulations.  The first case corresponds to the  setup of anisotropic inflation which can produce anisotropic hairs in cosmological background. The second case is the most favourable one from the view of \cite{Demozzi:2009fu}, because it does not suffer from the backreaction problem in the absence of stochastic effects.

\subsubsection{$n=3$}
\label{n3}
\begin{figure}[h!t!]
	\begin{subfigure}[b]{0.5\linewidth}
		\centering
		\includegraphics[scale=0.23]{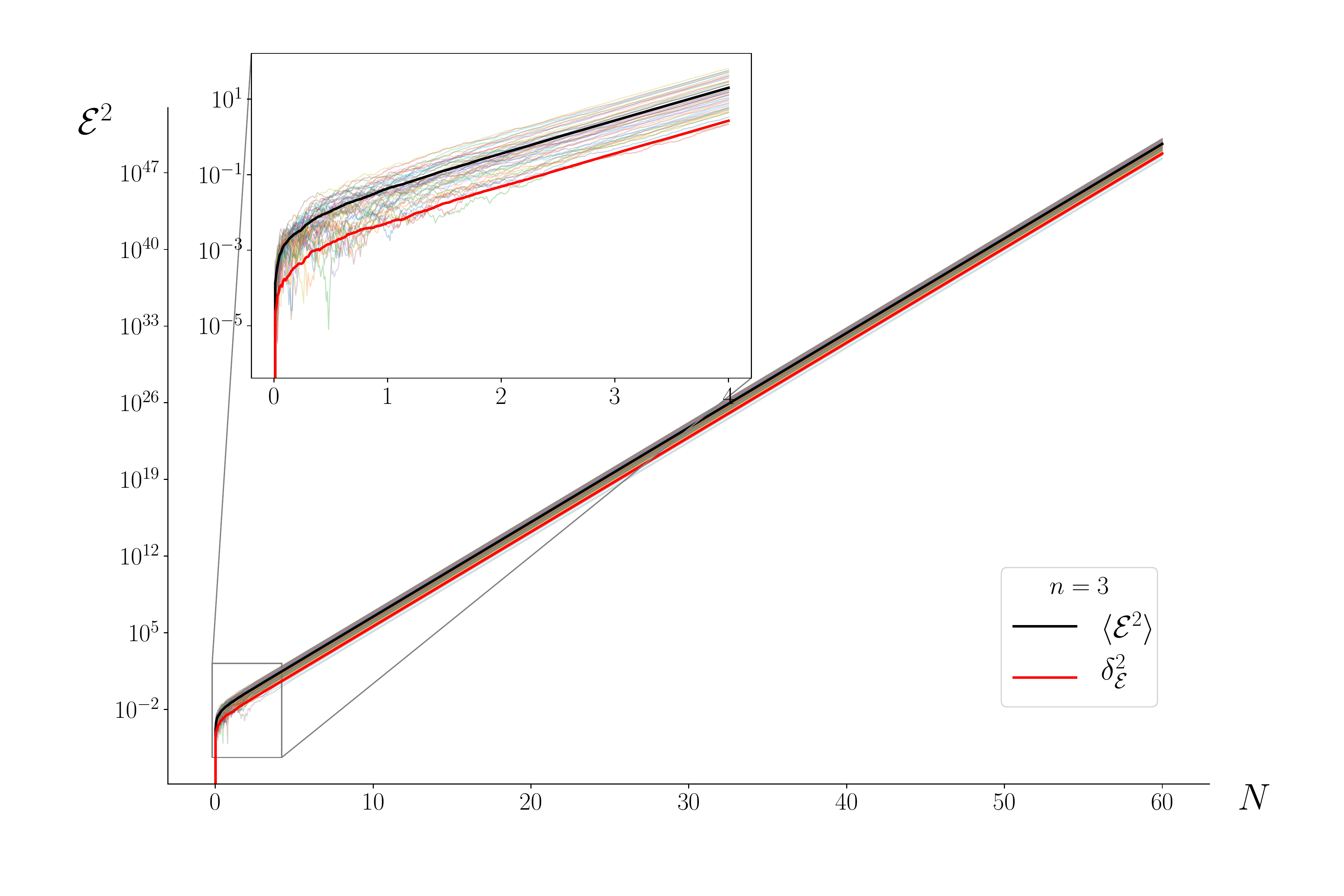}
		\caption{}
		\label{fig:En3}
		\vspace{4ex}
	\end{subfigure}
	\begin{subfigure}[b]{0.5\linewidth}
		\centering
		\includegraphics[scale=0.23]{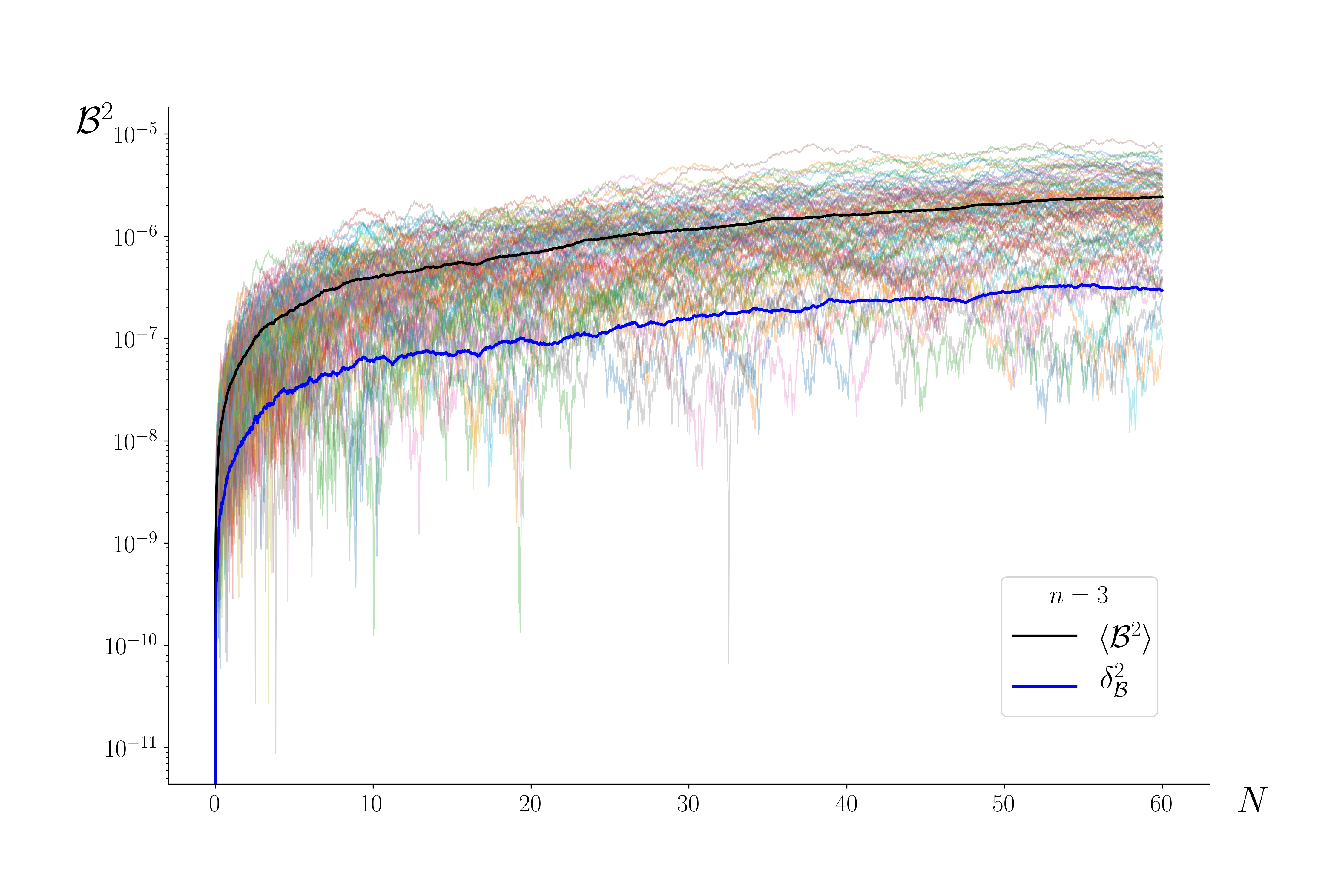}
		\caption{}
		\label{fig:Bn3}
		\vspace{4ex}
	\end{subfigure}
	\begin{subfigure}[b]{0.5\linewidth}
		\centering
		\includegraphics[scale=0.23]{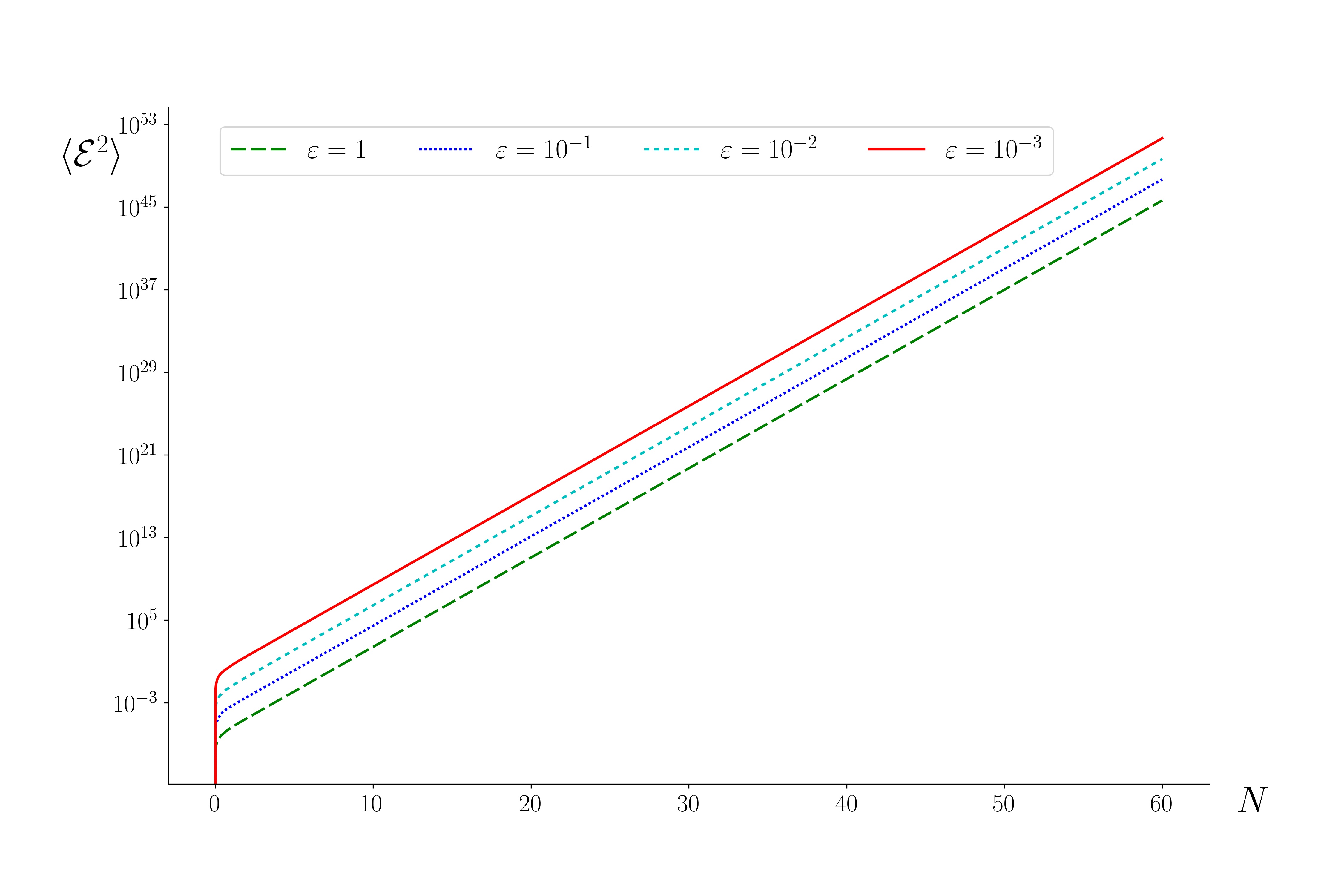}
		\caption{}
		\label{fig:epsn3}
	\end{subfigure}
	\begin{subfigure}[b]{0.5\linewidth}
		\centering
		\includegraphics[scale=0.23]{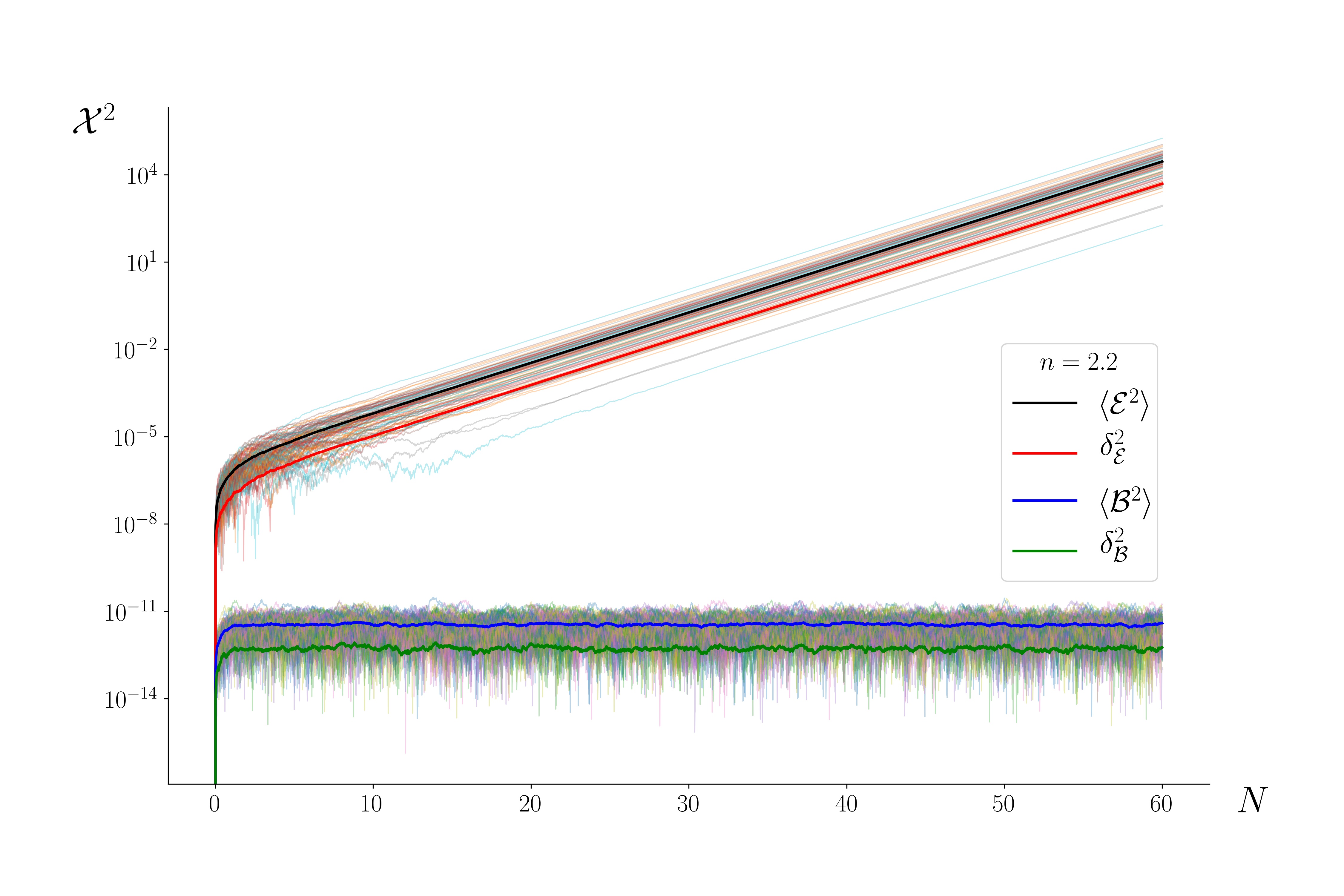}
		\caption{}
		\label{fig:Xn22}
	\end{subfigure}
	\caption{(a) and (b): Simulations of the evolution of $\langle {\cal X}^2 \rangle$ in terms of $N$ in weak coupling regime for $\varepsilon=10^{-3}$ for one hundred realization.~(a):  The growing behaviour of the electric fields.  As seen, at early stage  during about the first $4$ e-folds, the stochastic effects enhance the amplitude of the electric fields while the stochastic effects disappear after that.  (b):  The evolution of magnetic field for one hundred realizations in the case $n=3$. (c): The dependency of $\langle {\cal E}^2 \rangle$ to $\varepsilon$ for $n=3$.~(d): The evolution of electric and magnetic fields for one hundred realizations in the case $n=2.2$.}
	\label{fig:n3}
\end{figure}
The stochastic differential equations of magnetic field ($\nu=5/2$) and electric field ($\nu=7/2$) up to leading order are given by
\begin{eqnarray}
	{\rm d}\boldsymbol{\cal B} &=& -\epsilon_H~ \boldsymbol{\cal B}~{\rm d}N + \sqrt{6 {\cal P}_\zeta}~{\rm d}\boldsymbol{W}  \,, \label{B}
	\\
	{\rm d}\boldsymbol{\cal E} &=& \boldsymbol{\cal E}~ {\rm d}N+ \dfrac{35\sqrt{{\cal P}_\zeta}}{\sqrt{6}~\varepsilon}~{\rm d}\boldsymbol{W}  \,. \label{E}
\end{eqnarray}
Here, the backreaction from the electric energy density spoils the condition $R \ll 1$ very soon at $N_{\rm vio} \simeq 6$ e-folds, long before the magnetic field reaches to its stationary value which would be  given by Eqs.~\eqref{B2n-2} at around $N_{\rm eq} \simeq 2 \times 10^5$. This is the well-known backreaction problem of the electric fields in weak coupling regime \cite{Demozzi:2009fu}. 
The evolution of the means and the variances of the electric and magnetic fields for one hundred realizations till e-fold number $N=60$ are plotted in Fig.~\ref{fig:En3} and~\ref{fig:Bn3}. Also the dependency of $\langle {\cal E}^2 \rangle$ to $\varepsilon$ is shown in Fig.~\ref{fig:epsn3}.

\subsubsection{$n=2.2$}
\label{n2.2}
For $n=2.2$, the stochastic differential equations of magnetic field ($\nu=1.7$) and electric field ($\nu=2.7$)  up to leading order are given by
\begin{align}
	{\rm d}\boldsymbol{\cal B} &= -0.8~ \boldsymbol{\cal B}~{\rm d}N+ 1.14\sqrt{{\cal P}_\zeta}~\varepsilon^{4/5}~{\rm d}\boldsymbol{W}  \,, \label{B2.2}
	\\
	{\rm d}\boldsymbol{\cal E} &= +0.2~ \boldsymbol{\cal E}~ {\rm d}N+ 3.4\sqrt{{\cal P}_\zeta}~\varepsilon^{-1/5}~{\rm d}\boldsymbol{W}  \,. \label{E2.2}
\end{align}
Fig.~\ref{fig:Xn22} shows the evolution of the electric and magnetic field till 60 e-folds.

The condition $R\ll 1$ is violated and the backreaction of the electric field spoils inflation at around $N_{\rm vio} \simeq 55$ which is about the minimum  period of  inflation required.
The magnetic field reaches to its equilibrium   value
\begin{eqnarray}
	\langle \mathcal{B}^2 \rangle_{\rm eq} &\simeq&  5.4 \times 10^{-9} ~\varepsilon^{8/5} \,,
	\label{B22n3}
\end{eqnarray}
at around $N_{\rm eq} \simeq 3$, before the backreaction effects become important at $N_{\rm vio} \simeq 55 $.  Correspondingly, the present amplitude of  magnetic field from Eq.  \eqref{Bnow} is obtained to be 
\begin{align}
	B_{\rm now} &\simeq 1.9 \times 10^{-11}~G
	~\varepsilon^{4/5} \,.
\end{align}
Using the value of $\varepsilon$ at ${\rm Mpc}$ scales from Eq. \eqref{ec} this leads to 
\begin{align}
\label{B22sto}
	B_{\rm now} \simeq 4 \times 10^{-13}~ G \, .
\end{align}
This is  larger by 17 orders of magnitude compared to the result \eqref{B22standard} obtained in the conventional approach where the stochastic effects are neglected. 
Thus, taking the stochastic effects into account,   a model with the parameter $n \lesssim 2.2$ is promising to explain the origin of the primordial magnetic fields without being plagued by the electric field backreactions.  

\subsubsection{$n=2$}
\label{n2}
As mentioned before, this case corresponds to the setup of anisotropic inflation 
in which  observable anisotropic hairs can be generated during inflation \cite{Watanabe:2009ct,Emami:2013bk,Bartolo:2012sd}. 

The stochastic differential equations for the electromagnetic fields to leading order are given by
\begin{align}
{\rm d}\boldsymbol{\cal B} &= -\boldsymbol{\cal B}~ {\rm d}N+\dfrac{5\sqrt{{\cal P}_\zeta}}{2\sqrt{6}}~\varepsilon~{\rm d}\boldsymbol{W}  \,. \label{n2B}
\\
{\rm d}\boldsymbol{\cal E} &= -\epsilon_H~ \boldsymbol{\cal E} ~{\rm d}N+ \sqrt{6{\cal P}_\zeta}~{\rm d}\boldsymbol{W}  \,. \label{n2E}
\end{align}
This case is very desirable because the magnetic field reaches to its equilibrium value $\langle {\cal B}^2_{\rm eq} \rangle \simeq 2.2 \times 10^{-13}$ only after $N_{\rm eq} \simeq 3$ e-folds.   Thus, the observed magnetic field today on ${\rm Mpc}$ scale is given by
\begin{align}
B_{\rm now} &\simeq 1.2 \times 10^{-13}~G \, ,
\end{align}
which is acceptable for magnetogenesis. Interestingly, the backreaction effects are not important  in this case as we have $R \simeq 5 \times 10^{-10}$ after 60 e-folds.
This  is an interesting and unexpected   result in our work demonstrating the crucial effects of the stochastic dynamics which were neglected in previous works.

\subsubsection{Wiener processes in weak coupling regime}

According to subsection \ref{sec.b0}, the Wiener process for the magnetic field happens for  $n=3+\epsilon_H$.  In this case we obtain
\begin{align}
	{\rm d}\boldsymbol{\cal B} &= \sqrt{6 {\cal P}_\zeta}~ \varepsilon^{-\epsilon_H}~{\rm d}\boldsymbol{W}  \,. \label{n3b0}
	\\
	{\rm d}\boldsymbol{\cal E} &= \boldsymbol{\cal E} ~{\rm d}N+ \dfrac{35\sqrt{{\cal P}_\zeta}}{\sqrt{6}~\varepsilon}~{\rm d}\boldsymbol{W}  \,. \label{E3Wiener}
\end{align}
The strengths of the magnetic fields at the present time is given by 
Eq. \eqref{BWiener-now}. But as in the case of $n=3$ in strong coupling regime, the backreactions kick in at about  $N_{\rm vio}^W\simeq 6$ and inflation is terminated quickly.  Thus this case also does not work because of  the electric field backreactions.

On the other hand, a Wiener process for the electric field occurs for $n=2+\epsilon_H$ in which
\begin{align}
	{\rm d}\boldsymbol{\cal E} &= \sqrt{6 {\cal P}_\zeta} \varepsilon^{-\epsilon_H}~{\rm d}\boldsymbol{W}  \,. \label{n2b0E}
\end{align}
However, the dynamics of the magnetic field in this case is governed by an OU process with the  amplitude of the present time magnetic fields similar to the case of $n=2$.

\subsection{Summary of the results}
\label{sum}

\begin{figure}[t!]
	\begin{subfigure}[b]{0.5\linewidth}
		\centering
		\includegraphics[scale=0.235]{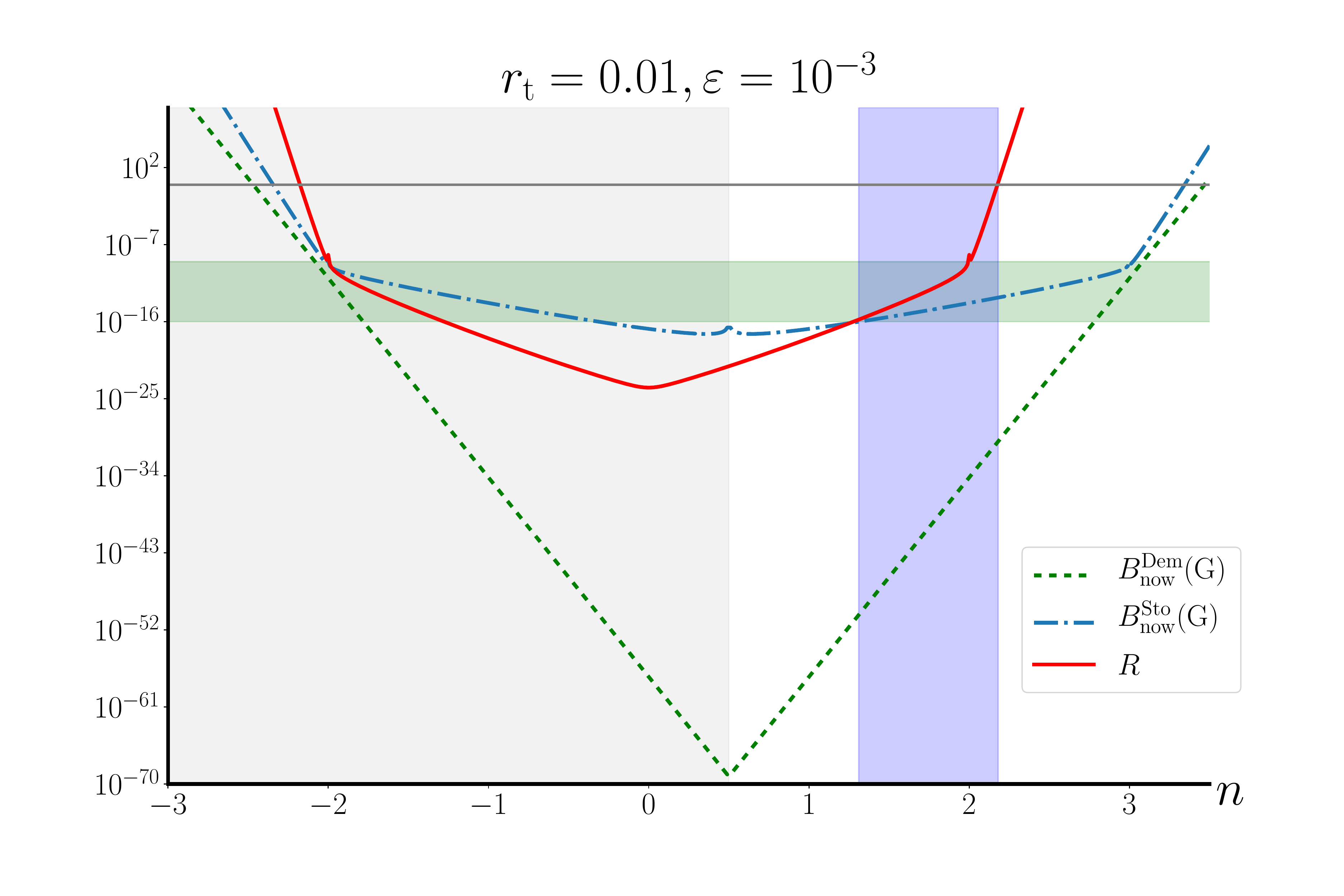}
		\caption{}
		\label{fig:BsBqrt-2ep10-3}
		\vspace{2ex}
	\end{subfigure}
	\begin{subfigure}[b]{0.5\linewidth}
		\centering
		\includegraphics[scale=0.235]{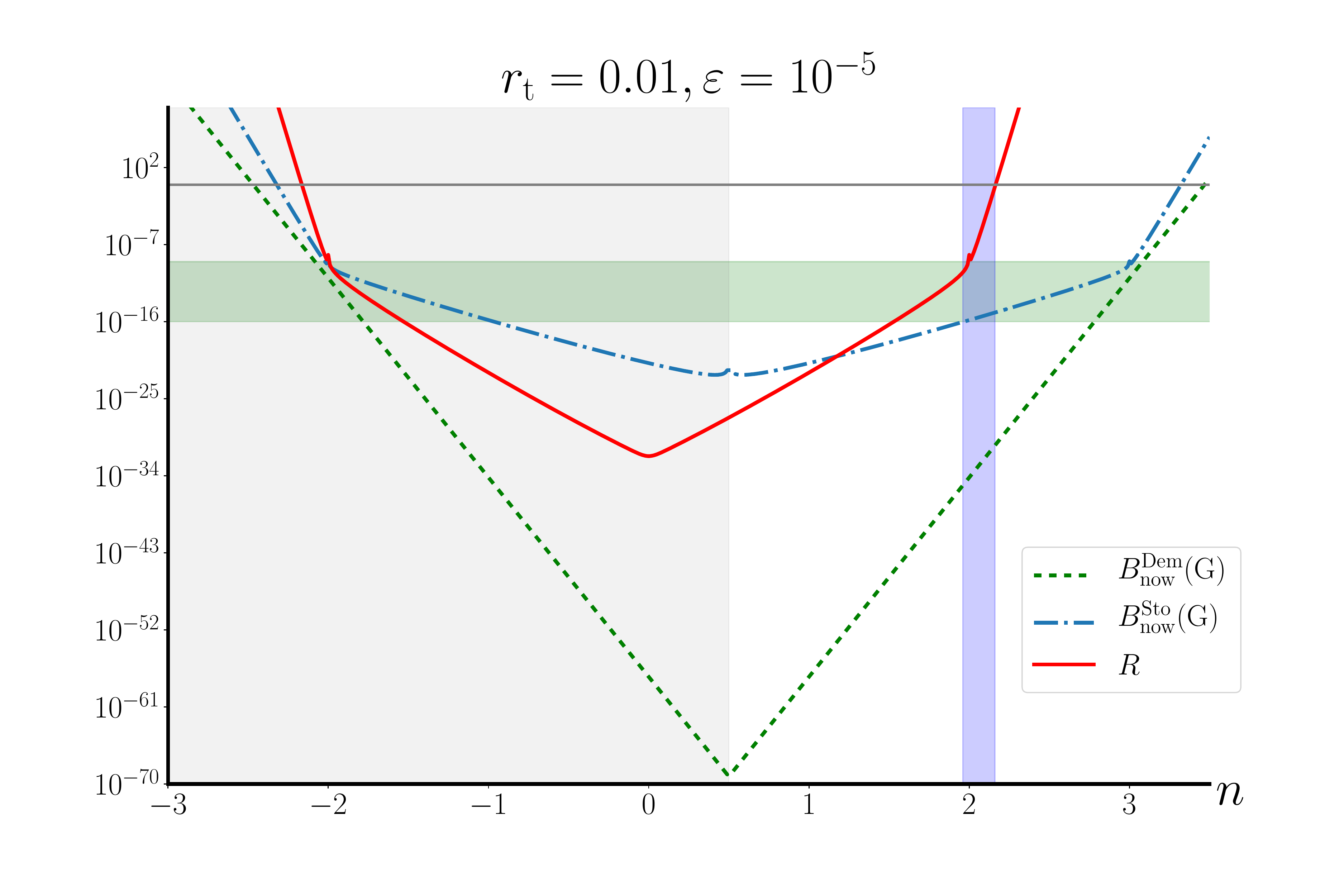}
		\caption{}
		\label{fig:BsBqrt-2ep10-5}
		\vspace{2ex}
	\end{subfigure}
	\begin{subfigure}[b]{0.5\linewidth}
		\centering
		\includegraphics[scale=0.235]{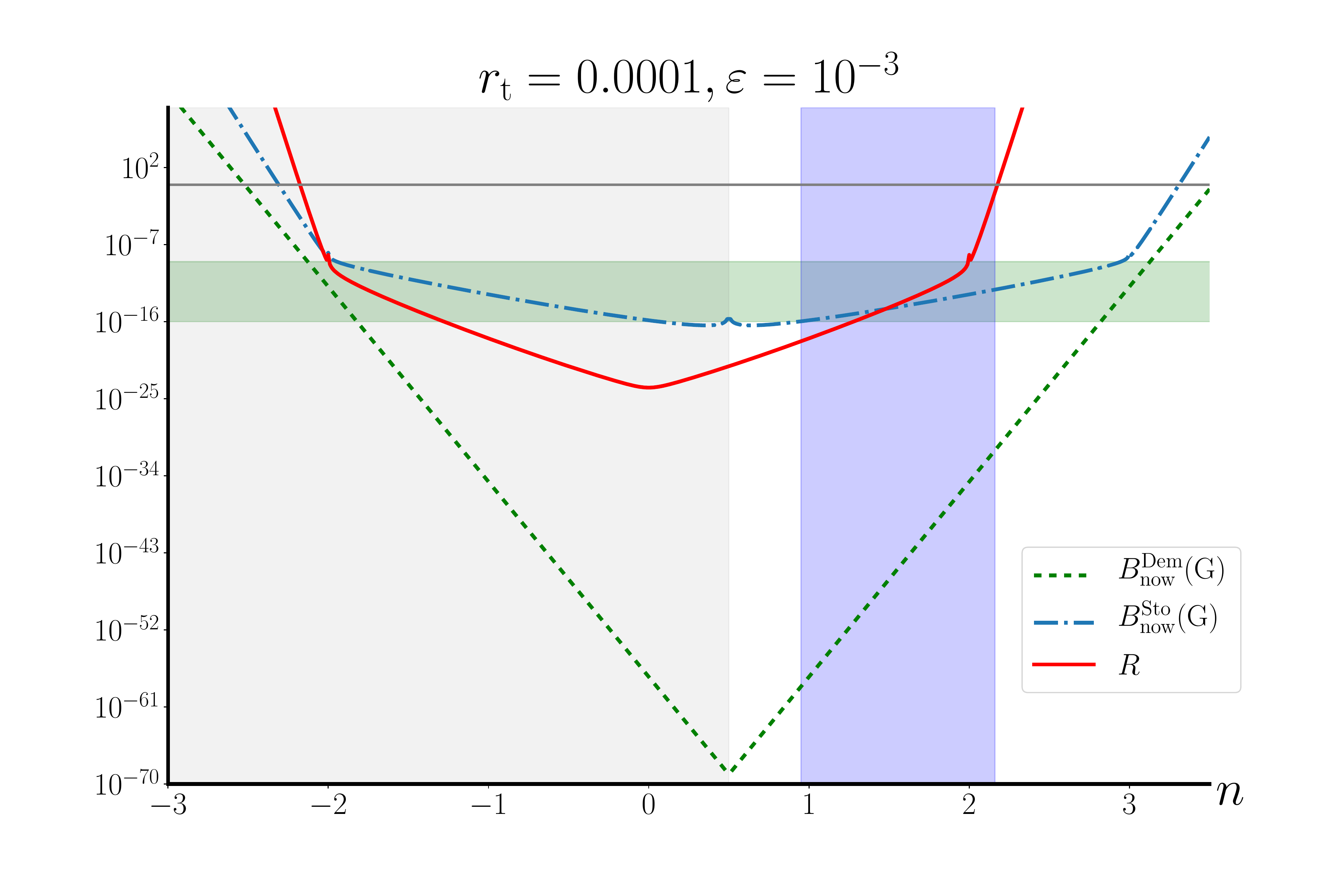}
		\caption{}
		\label{fig:BsBqrt-4ep10-3}
	\end{subfigure}
	\begin{subfigure}[b]{0.5\linewidth}
		\centering
		\includegraphics[scale=0.235]{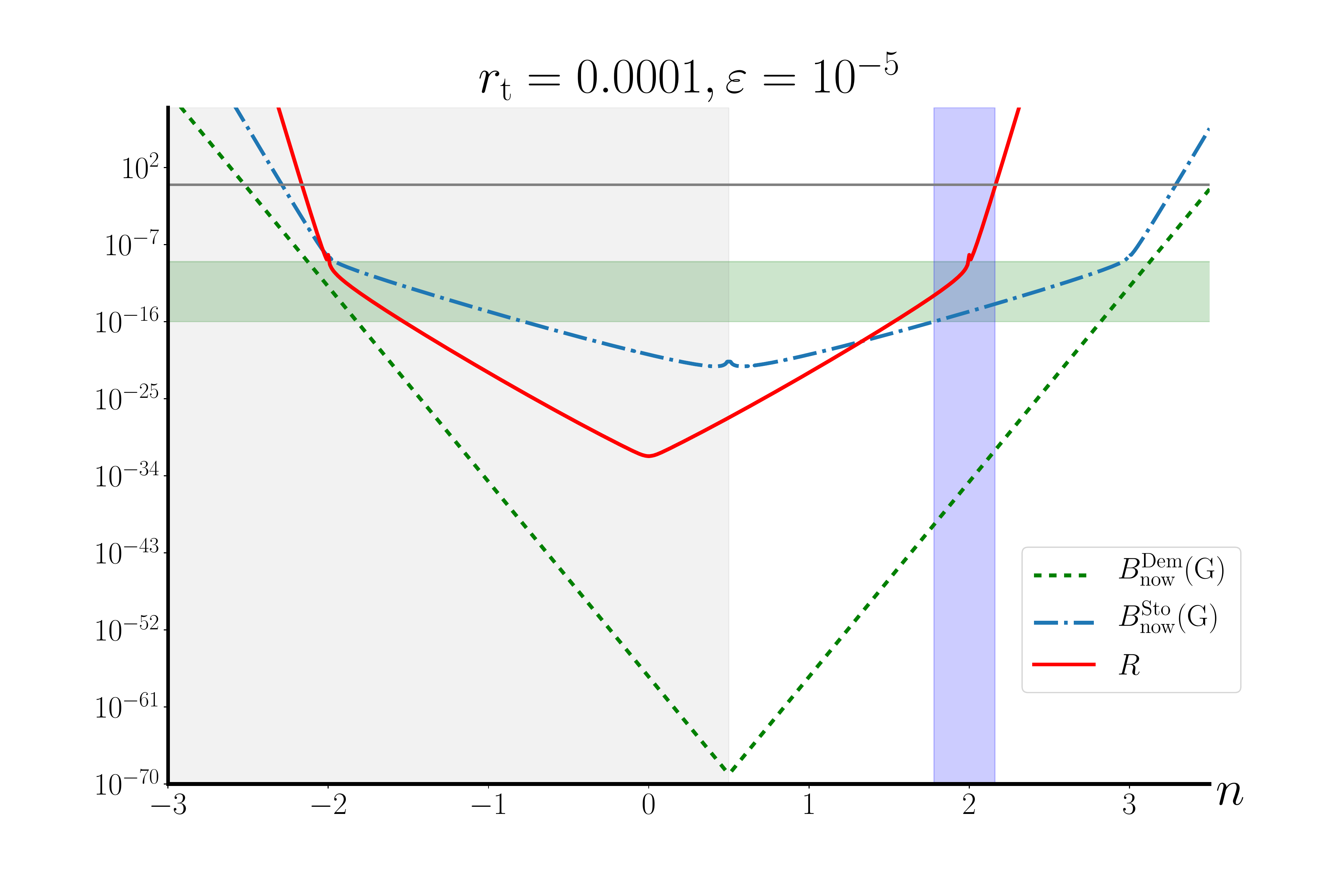}
		\caption{}
		\label{fig:BsBqrt-4ep10-5}
	\end{subfigure}
	\caption{Comparison of the amplitudes of the present day magnetic fields   between usual approach employed by Demozzi \textit{et al.} \cite{Demozzi:2009fu} $(B_{\rm now}^{\rm Dem})$ and the stochastic approach $(B_{\rm Now}^{\rm Sto})$. 
The vertical axis shows simultaneously the values of  $B_{\rm now}^{\rm Dem}$ and $B_{\rm Now}^{\rm Sto}$ in units of Gauss  and the value of $R$  with
 $R=1$ indicated by the horizontal solid line.  The grey region corresponds to the strong coupling regime while elsewhere represents the weak coupling regime. We have assumed that inflation last about $60$ e-folds. The green horizontal band is the observational bound \eqref{B-bound}. The blue vertical band shown in the weak coupling regime indicates a healthy parameter space for $n$ in which the model has not been plagued by the backreaction effects. As seen, the blue band becomes wider if $r_{\rm t}$ decreases or $\varepsilon$ increases.}
	\label{fig:BsBq}
\end{figure}
\begin{figure}[t]
	\begin{subfigure}{0.5\linewidth}
		\centering
		\includegraphics[scale=0.241]{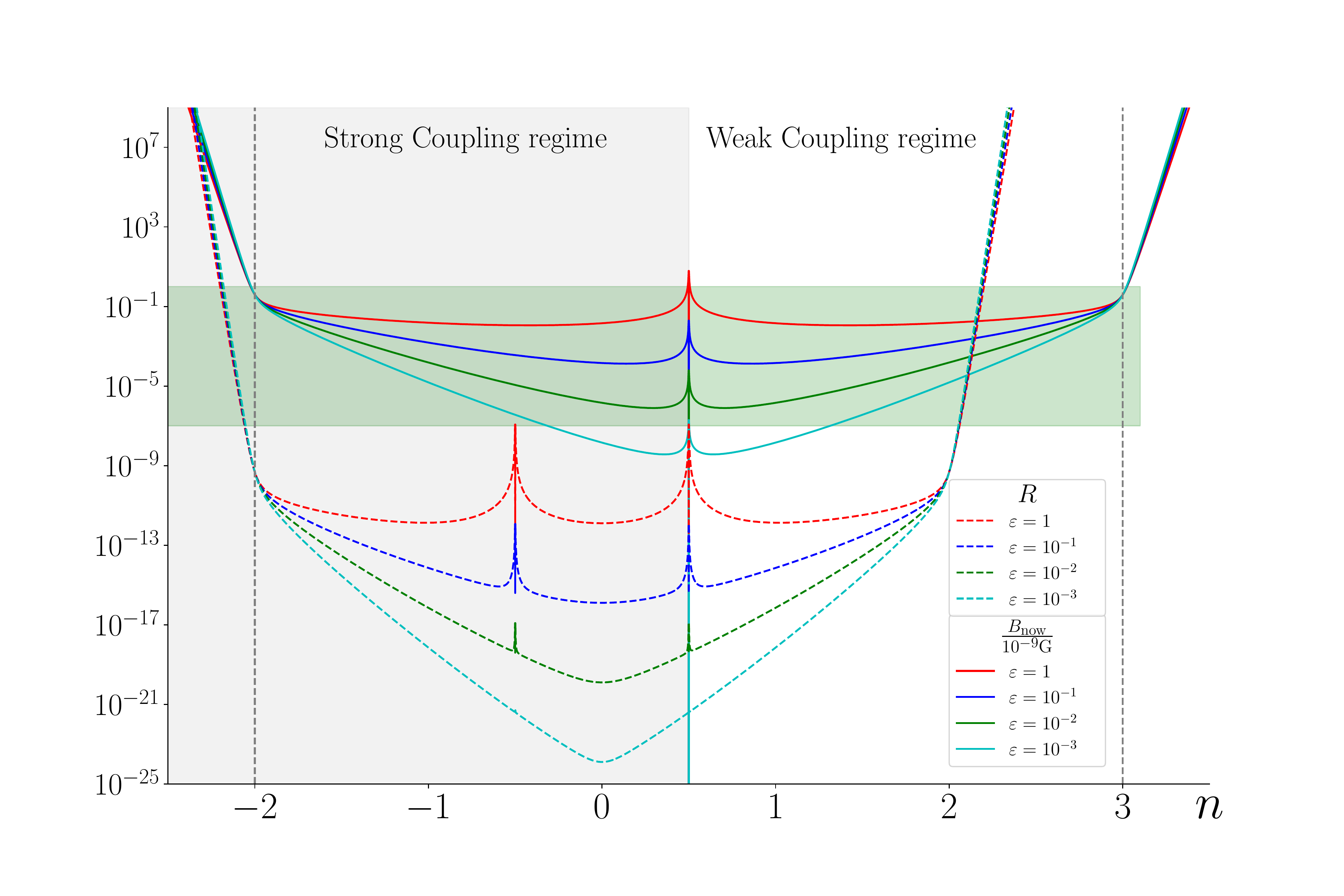}
		\caption{}
		\label{fig:RBnow}
	\end{subfigure}
	\begin{subfigure}{0.5\linewidth}
		\centering
		\includegraphics[scale=0.45]{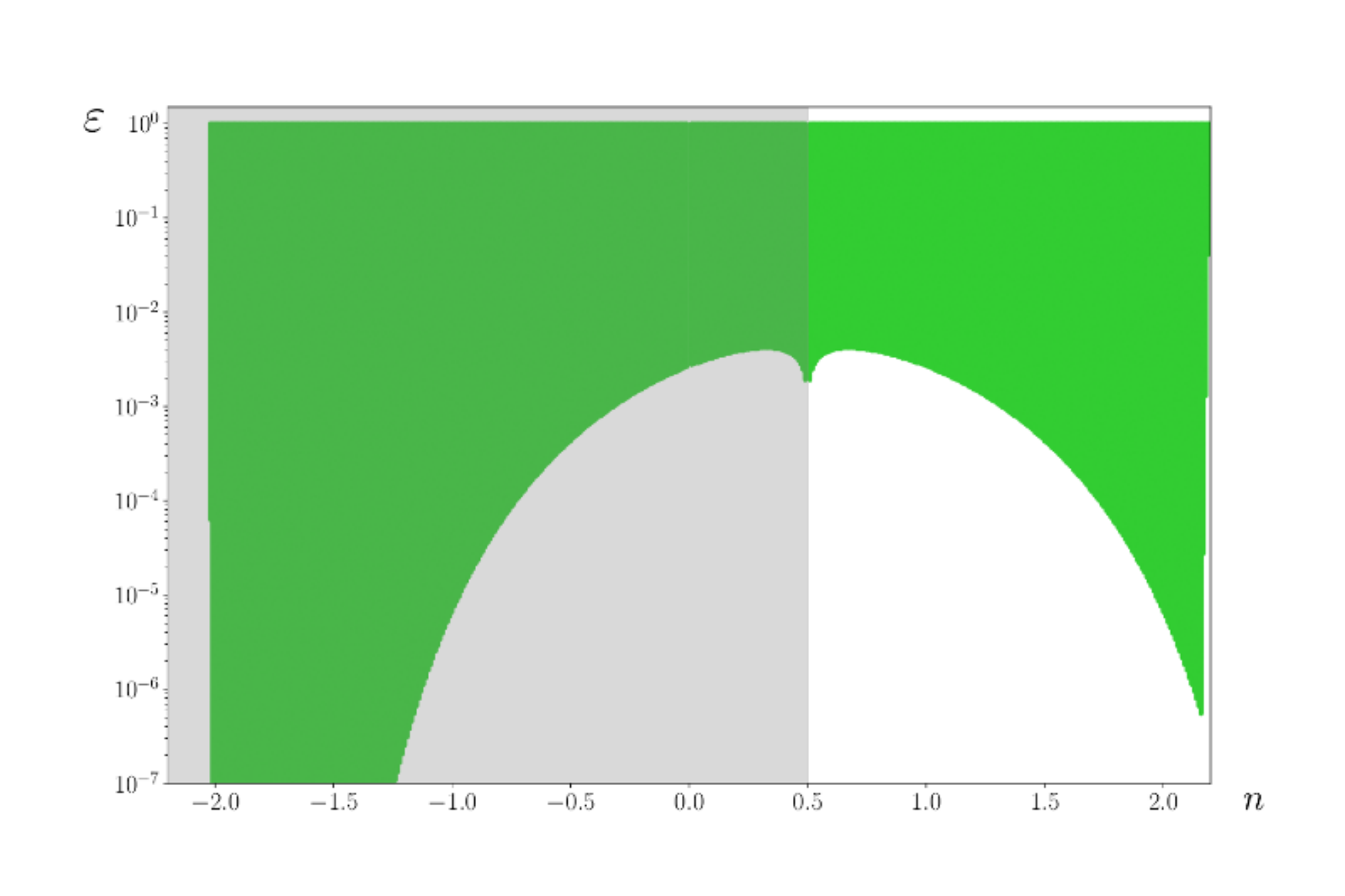}
		\caption{}
		\label{fig:GreenBound}
	\end{subfigure}
	\caption{(a): A snapshot of $R$ at the moment $N=60$ e-fold is given by the dashed curves while the solid curves indicate the present value of the magnetic field (normalized to femto Gauss) assuming  inflation lasts $60$ e-folds with $r_{\rm t}=0.01$.  The vertical axis shows simultaneously the amplitudes of $R$ and the normalized magnetic field. The green region corresponds to the observational bound \eqref{B-bound}. Above the green bound we have  $R > 1$. The vertical dashed lines at $n=-2$ and $n=3$ represents the scale invariant spectrum for the magnetic field. (b): The green regions are  allowed  in the parameter space of $n$ corresponding to the bound \eqref{B-bound} intersected by the condition  $R <1$. 
	}
\label{fig:RBnowGreenBound}
\end{figure}

In this subsection we present our results for generating primordial magnetic fields in the  $f^2 F^2$ model without encountering the backreaction or strong coupling problems. We present some plots to show that there is a suitable range of parameter space which results in desired observable magnetic fields in the range \eqref{B-bound}.

Before presenting the plots, let us take a closer look at the evolution of the magnetic field for the case $n=2$ and compare the results in the presence and in the absence of the stochastic noises. As it is seen from Eq. \eqref{B_superhorizon}, in the absence of stochastic effects the magnetic field rapidly decreases during inflation. But in the presence of stochastic noises, the stochastic differential equation \eqref{n2B} determines its evolution. Consequently, the magnetic  field falls into its stationary state at around $N_{\rm eq}=3$ e-folds. This clearly shows how stochastic effects change the fate of the magnetic fields. The results for the  magnetic fields from the conventional approaches presented  in Sec.~\ref{Ratra-today} based on the analysis of \cite{Demozzi:2009fu} and the stochastic approach are illustrated in Fig.~\ref{fig:BsBq} for the cases $r_{\rm t}=10^{-2}, 10^{-4} ({\rm GUT~inflation})$ and $\varepsilon=10^{-3}, 10^{-5}$. The former is denoted by $B_{\rm now}^{\rm Dem}$ while the latter is shown by $B_{\rm now}^{\rm Sto}$. Requiring that inflation lasts at least $N=60$ e-fold, Fig.~\eqref{fig:BsBq} indicates that the stochastic kicks enhance the strength of  magnetic field by many orders of magnitudes; in the weak coupling regime the ratio $B_{\rm now}^{\rm Sto}/B_{\rm now}^{\rm Dem}$ ranges in the interval $  \sim 10^{2} - 10^{55}$~! The vertical blue band corresponds to the intersection of the green band Eq. \eqref{B-bound} and $R<1$. Therefore, the blue band is the allowed region for the parameter space of $n$ in which $B_{\rm now}^{\rm Sto}$ falls into the bound \eqref{B-bound} without being plagued by the backreaction problem. Above the green region and to the right of blue band 
we have  $R >1$ which for all cases the curve of $R$ leaves the blue band at around $n=2.1$. On the other hand, for the fixed  $\varepsilon=10^{-3}$, at around $n=1.2$ 
the curve of $R$ enters to the blue band. Therefore, the interval $n \in [1.2, 2.1]$ can be considered as a  good range for the parameter $n$ in which  the model works. The strength of generated magnetic fields in this interval  is around $\sim 10^{-16}-10^{-13}$ G. But note that the generated magnetic field is not scale invariant in this range. Also  note that by decreasing $r_{\rm t}$ or increasing $\varepsilon$ the blue band becomes wider.

In Fig.~\ref{fig:RBnow}, we have reproduced the results of Fig.~\eqref{fig:BsBq} for the cases $\varepsilon=1,0.1,0.01$ and $0.001$.  Fig.~\ref{fig:GreenBound} represents the allowed value of the parameter $n$  corresponding to the green band $10^{-16}{\rm G} < B_{\rm now} < 10^{-9}{\rm G}$ intersected by the condition  $R < 1$ for a wide range of $\varepsilon$. As can be seen,  for $\varepsilon \lesssim 5 \times 10^{-7}$ (corresponding to the coherent scale $\lambda_{\rm B} \lesssim 63 \,  {\rm pc}$), there is no allowed value of $n>1/2$ that the model can work. 

The conclusion is that, taking the stochastic effects into account,   there are some ranges of   parameter space in which the observed primordial magnetic field can be achieved without facing the backreaction or the strong coupling problems.

\section{Probabilistic analysis}
\label{probability}

In this section we briefly study  the probability distribution function of the magnetic field.
The probability distribution functions for the cases $b_\nu=0$ and $b_\nu < 0$ are given by 
Eqs. \eqref{f_X_b0} and \eqref{f_X_station} respectively. The former represents a Wiener process while the latter represents an OU process. 
These distribution functions enable us to calculate the probability of having a given amplitude for the magnetic field in a desired range, as presented in Eqs. \eqref{Pb0} and \eqref{Pbminus}. The desired range corresponds to the lower and upper bounds on cosmological magnetic fields as given  in Eq. \eqref{B-bound}. Subsequently, 
 these bounds are translated  into Eq. \eqref{B-end-bound} which determines the values of $\chi_1$ and $\chi_2$ in Eqs. \eqref{Pb0} and \eqref{Pbminus}.

\begin{figure}[h]
	\begin{subfigure}{0.5\linewidth}
		\centering
		\includegraphics[scale=0.235]{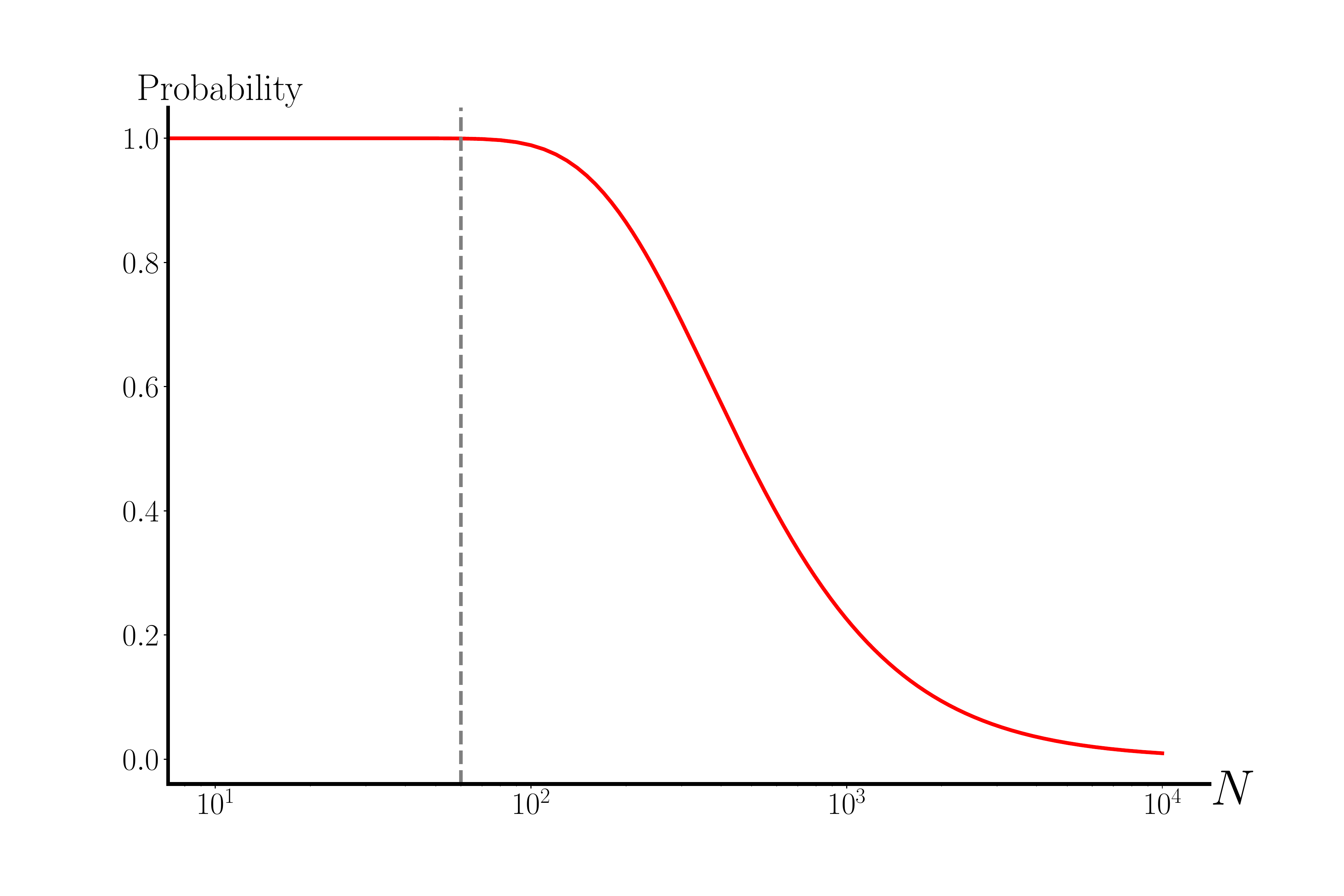}
		\caption{}
		\label{fig:Pb0}
	\end{subfigure}
	\begin{subfigure}{0.5\linewidth}
		\centering
		\includegraphics[scale=0.235]{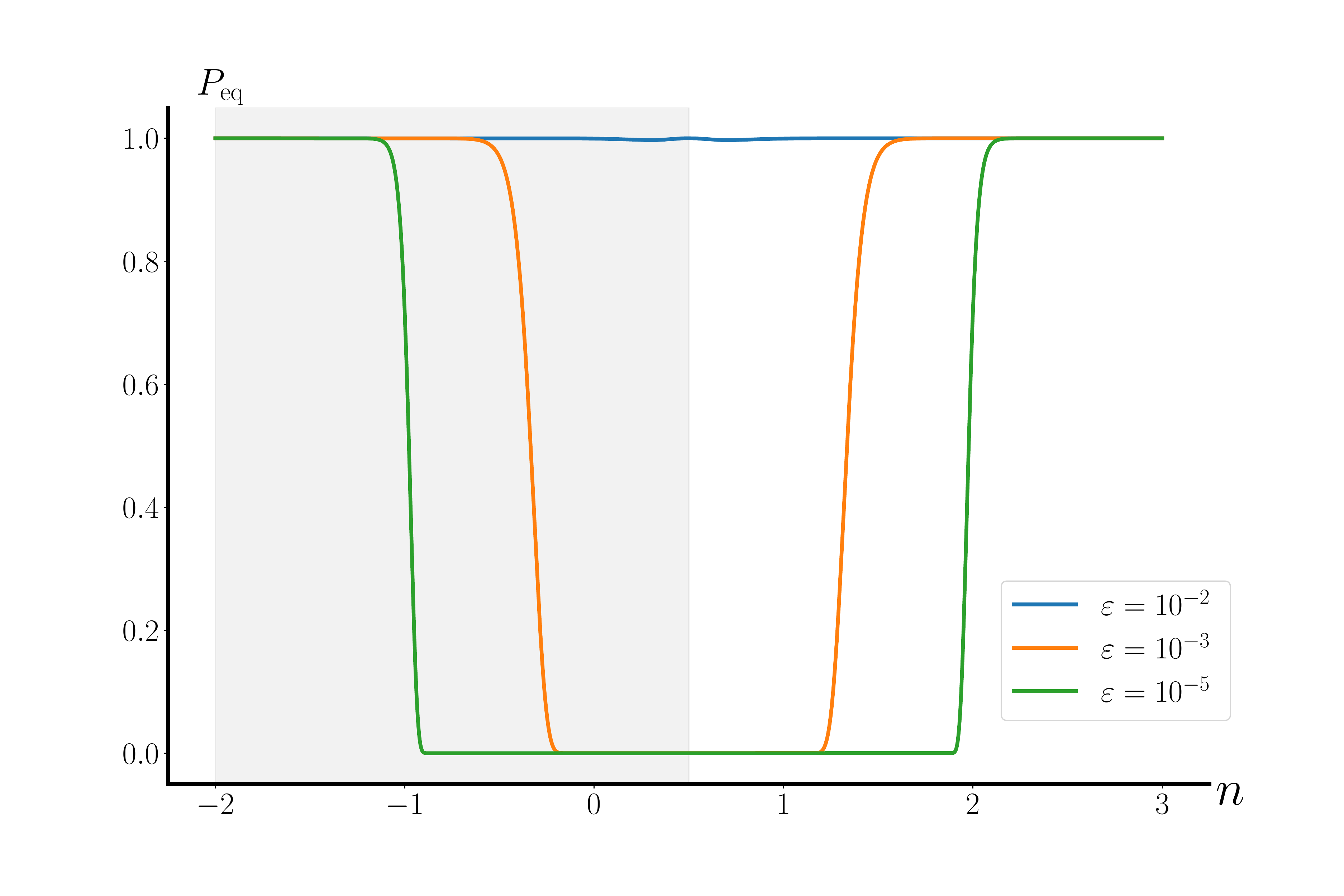}
		\caption{}
		\label{fig:Pbminus}
	\end{subfigure}
	\caption{(a) The probability of having the magnetic field in the range of \eqref{B-bound} according to Eq.~\eqref{Pb0} when  $b_\nu=0$ corresponding to
	$n=3+\epsilon_H $ and $n= -2-\epsilon_H$~. 
	The grey dashed line corresponds to  $N=60$ e-fold. The overall behaviour of this probability is independent of $\varepsilon$ but here the plot is for $\varepsilon=10^{-3}$. (b) The probability distribution Eq.~\eqref{Pbminus} in terms of parameter $n$ for the case with 
	$b_\nu <0 $ where magnetic fields reach to a equilibrium state.  The grey (white) region represents the strong (weak) coupling regime.  The probability depends on $\varepsilon$. For  $\varepsilon$ in the range $10^{-2} \lesssim \varepsilon \lesssim 1$, the probability of having the present value of magnetic field in the bound \eqref{B-bound} is almost $100\%$. This plot  is well consistent  with Fig.~\ref{fig:BsBq}}
	\label{fig:Probab}
\end{figure}

As mentioned in subsection \ref{sec.b0},  the case $b_\nu=0$ corresponds to  $|\nu|=\epsilon_H+5/2$. For the magnetic field there is two possibilities for the parameter  $n$: $n=3+\epsilon_H$ and $n=-2-\epsilon_H$~. For these two values, the probability of having the present magnetic field in the range of \eqref{B-bound} is given by  \eqref{Pb0},
\begin{align}
\label{Pb0greenbound}
P(10^{-10} \lesssim {\cal B}_{\rm end} \lesssim 10^{-3} ;N) \,.
\end{align}
The dependency of the probability to $N$ is due to the Wiener process which describes a random walk with variance proportional to $N$. Hence, the probability of having the magnetic field in a desired range depends on how many e-folds inflation was in progress. The evolution of this probability is plotted in Fig.~\ref{fig:Pb0}. The plot shows that if inflation takes about $100$ e-folds or less, then the probability is about $100{\%}$~.  But if the inflationary stage takes much longer period 
then the probability becomes smaller and smaller. The overall behaviour of the probability \eqref{Pb0greenbound} is independent of $\varepsilon$.

If $b_{\nu}<0$ (represented by the blue dashed lines in Fig.~\ref{fig:Neqbminus}), the magnetic field admits a stationary state at around the e-folding number $N_{\rm eq}$ defined in Eq. \eqref{Neq}. Here we have assumed that the magnetic field reaches to its equilibrium (stationary) state before the end of inflation i.e.  $N_{\rm eq} \lesssim 60$.
In this case the probability of having the present value of magnetic field in the interval determined in Eq.  \eqref{B-bound} is given by \eqref{Pbminus}, yielding
\begin{align}
P_{\rm eq}(10^{-10} \lesssim {\cal B}_{\rm end} \lesssim 10^{-3}) \,.
\end{align}
This probability depends on $\varepsilon$ and $n$ as it is shown in Fig.~\ref{fig:Pbminus}. The main feature in this plot is that  the smaller values of the parameter $\varepsilon$  yield smaller   probability in some parts of parameter space $n$. This statement is consistent with the results in Fig.~\ref{fig:BsBq} as the probabilistic interpretation based on the Fokker-Planck equation is a parallel approach to the mechanism of stochastic differential equations presented in section \ref{BToday}.
\section{Conclusion}
\label{Conclusion}

In this paper, we have revisited the mechanism of magnetogenesis in the $f^2 F^2$ inflationary model by taking into account the stochastic effects.  We have  derived the associated Langevin equations for the electric and magnetic fields. We have also derived and solved the Fokker-Planck equation associated  to the magnetic field for wide ranges of parameter  $n$.  

As mentioned in previous literature, there are difficulties in magnetogenesis in the $f^2 F^2$ model,
namely the backreaction and the strong coupling problems. If one chooses to work in the weak coupling  regime the backreaction of the electric fields would spoil inflation too early. In usual approach employed in \cite{Demozzi:2009fu},  there is region  of the parameter space $n$ ($n<2.2$) in which the backreactions are not strong to spoil  inflation, but the amplitude of the generated magnetic field cannot exceed $10^{-30}{\rm G}$ in Mpc scales today (for example, for the most favourable case $n=2.2$, see Eq. \eqref{B22standard}). Although in the presence of the stochastic noises the backreaction of the electric field becomes more relevant to spoil inflation, but at the same time these stochastic effects enhance the magnetic field significantly. For example, for the case of $n=2.2$, the stochastic effects amplify the magnetic fields by about 17 orders of magnitude, see Eq. \eqref{B22sto}. The main reason for the amplification of the magnetic field in this case (and similar cases with $b_\nu<0$)  is  due to the \textit{mean-reverting} process of an Ornstein-Uhlenbeck stochastic differential equation which settles the fields into an equilibrium state and prevent them from decaying. This is unlike the conventional approach in which the magnetic fields  decay rapidly on superhorizon scales.

In Sec. \ref{BToday} we have shown that for $n $ in the range $n \in [1.2,2.1]$  magnetic fields at the coherent length ${\rm Mpc}$ with amplitude $\sim 10^{-16}-10^{-13} {\rm G}$ at present time can be generated  without encountering the backreaction or the strong coupling problem (the blue band in Fig.~\ref{fig:BsBq}). This range of allowed value of $n$ actually depends on $r_{\rm t}$ and $\varepsilon$ in which  the lower bound on the allowed value of $n$ can be pushed   to 0.51.  In addition, the probabilistic analysis performed in Sec.~\ref{probability} confirmed this conclusion. 

To have a consistent model, one should also check that the  curvature perturbation induced by the electromagnetic fields is consistent with the CMB observations. In \cite{Talebian:2019opf} we have studied  the case $n\simeq2$ by allowing a background electric field energy density, ${\cal E}_{\rm cl} \neq 0$. There, we have shown that the theory is consistent with the CMB observations (i.e. the probability of generating quadrupolar statistical anisotropy is consistent with CMB constraints). Combining our finding in \cite{Talebian:2019opf} with our current results show that the anisotropic inflation model $f^2 F^2$ with $f \propto a^{-2}$ (i.e. $n=2$) is not only consistent with the CMB anisotropy constraints but also can produce primordial seeds for magnetic fields observed today on Mpc scales. This is one important result of this work.

Another important finding in this work is about the roles of the stochastic effects in  backreaction problem. In strong coupling regime  Demozzi \textit{et al.} \cite{Demozzi:2009fu} have shown that the magnetic field energy density does not spoil inflation if $n \ge -2.2$. However,  we have seen  that in the presence of the stochastic noises the backreactions of the magnetic field can spoil inflation even for this range of $n$ (e.g. refer to subsection \ref{n-2.2}). Moreover,  the backreaction problem in the weak coupling regime becomes more important in the presence of stochastic noises than in their  absence if $n>2.2$ (e.g. refer to the subsection \ref{n3} and Fig.\ref{fig:n3}).

This study opens up a new window into the studies of the primordial magnetogenesis 
in the presence of stochastic noises. For example,  it would be interesting to explore the stochastic approach in other scenarios such as magnetogenesis in the setup of pseudo scalar inflation \cite{Anber:2006xt, Caprini:2014mja}. Also one can look at
 Schwinger effects in this setup \cite{Sobol:2019xls, Shakeri:2019mnt}, Faraday's law of induction \cite{Kobayashi:2019uqs} and conductivity \cite{Fujita:2019pmi}  in the presence of the stochastic noises. A good question is how stochastic noises affect the parametric resonance mechanism to produced magnetic fields \cite{Patel:2019isj}.  We leave these issues to future works.

\vspace{0.5cm}

 {\bf Acknowledgments:}  A. N. would like to thank Tokyo Institute for Technology (TITECH) for hospitality while this work was in progress.

\appendix

\section{Properties of the noise terms $(\boldsymbol{\tau},\boldsymbol{\sigma})$}
In this appendix, we derive the explicit forms of the quantum noises $(\boldsymbol{\tau},\boldsymbol{\sigma})$
and then obtain the associated Langevin equation for the superhorizon perturbations. 

The noise terms are given by
\label{noise}
\begin{align}
\boldsymbol{\tau}(t,\boldsymbol{x}) = \varepsilon a H^2 \int \frac{d^3k} {\left(2\pi\right)^3} \, \delta \left(k-\varepsilon aH\right) \boldsymbol{\dot{X}}_{\boldsymbol{k}}(t)~ e^{i\boldsymbol{k}.\boldsymbol{x}} \,,\label{tau_Xdot-app}
\\
\boldsymbol{\sigma}(t,\boldsymbol{x}) = \varepsilon a H^2 \int \frac{d^3k} {\left(2\pi\right)^3} \, \delta \left(k-\varepsilon aH\right) \boldsymbol{X}_{\boldsymbol{k}}(t)~ e^{i\boldsymbol{k}.\boldsymbol{x}} \,,\label{sigma_X-app}
\end{align}
with the  mode function
\begin{align}
X_\lambda &= i\dfrac{\sqrt{\pi}}{2}~k H^2 ~\eta^{5/2} ~H^{(1)}_{\nu}(-k\eta) \,,
\label{X_mode-app}
\end{align}
appearing in its associated Fourier transforms
\begin{eqnarray}
\boldsymbol{X}_{\boldsymbol{k}}(\eta) &=& \sum_{\lambda = \pm} \boldsymbol{e}^\lambda(\hat{\boldsymbol{k}}) \left[ X_\lambda(\eta,k)\,\hat{a}^\lambda_{\boldsymbol{k}} +  X_{\lambda}^{*}(\eta,k)\, \hat{a}^{\lambda \dagger}_{-\boldsymbol{k}} \right] \, .
\label{X_lambda-app}
\end{eqnarray}

The quantum properties of the noises $(\boldsymbol{\tau},\boldsymbol{\sigma})$
can be read off by looking at their commutators. 
Here,  to simplify the notation, 
we denote the components of the source terms $(\boldsymbol{\tau},\boldsymbol{\sigma})$ collectively  as $A_i$ and $B_i$ and define their commutators  as 
\begin{eqnarray}
\dfrac{1}{2}\left[ A_i(x_1),B_j(x_2) \right]
\equiv
D_{AB}(x_1,x_2)~j_0\big( \varepsilon aH |\bfx_1-\bfx_2| \big) ~\delta_{ij} ~\delta(t_1-t_2)\,,
\label{commutator}
\end{eqnarray}
in which $x_i=(t_i,\boldsymbol{x}_i)$ and $j_0$ is the zeroth order Bessel function.
Since $\varepsilon$ is a small parameter, and using the definitions of \eqref{X_mode}, \eqref{sigma_X} and \eqref{tau_Xdot},  we obtain
\begin{align}
\label{correlation}
D_{\sigma \sigma}(x_1,x_2) &= D_{\tau \tau}(x_1,x_2) =0 \,,
\\
\label{correlation2}
D_{\sigma \tau}(x_1,x_2) &= -i \frac{H^6}{6\pi^2}~\varepsilon^{5} \,.
\end{align}
As we see, the quantum non-commutativity disappears up to ${\cal O}(\varepsilon^4)$ independent of $\nu$. Hence, by choosing a sufficiently small value of $\varepsilon$, the quantum nature of these noises dies away and one can treat them as classical noises.

The correlation functions of $(\boldsymbol{\tau},\boldsymbol{\sigma})$ can be defined by choosing a suitable state. Here we impose the Bunch-Davies (Minkowski) initial condition $|0 \rangle$ and define the correlation function among operators $A_i$ and $B_j$ as
\begin{eqnarray}
\left \langle0 \left| A_i (x_1) ~ B_j (x_2) \right| 0\right\rangle
\equiv
C_{AB}(x_1,x_2)~j_0\big( \varepsilon aH |\bfx_1-\bfx_2| \big) ~\delta_{ij} ~\delta(t_1-t_2)
\label{correlator}
\,.
\end{eqnarray}
 Consequently,  for $\nu \neq 0$ we obtain
\begin{align}
C_{\sigma \sigma}(x_1,x_2)
&\approx
\dfrac{4^{-1+|\nu|}\left(\Gamma\left(|\nu|\right)\right)^2}{ 3\pi^3}~H^5~\varepsilon^{5-2|\nu|} ,
\label{ss}
\\
\nonumber\\
C_{\tau \tau}(x_1,x_2)
&\approx
\left\lbrace \begin{array}{lc}
\dfrac{H^7~\varepsilon^4}{ 6\pi^2} &\nu = \pm 5/2\\
\\
\dfrac{4^{-1+|\nu|}\left(\Gamma\left(|\nu|\right)\right)^2}{ 3\pi^3}~H^7~\varepsilon^{5-2|\nu|}(\dfrac{5}{2}-|\nu|)^2 & \nu \neq \pm 5/2 , \\
\\
\end{array}\right.
\label{tt}
\\
C_{\sigma\tau+\tau\sigma}(x_1,x_2)
&\approx
\left\lbrace \begin{array}{lc}
\dfrac{H^6~\varepsilon^2}{\pi^2} &\nu = \pm 5/2\\
\\
\dfrac{4^{-1+|\nu|}\left(\Gamma\left(|\nu|\right)\right)^2}{ 3\pi^3}~H^6~\varepsilon^{5-2|\nu|}(5-2|\nu|) & \nu \neq \pm 5/2 . \\
\\
\end{array}\right.
\label{st}
\end{align}
However, for $\nu = 0$, the leading order  correlation function  starts at the order 
$\sim {\cal O}(\varepsilon^5)$ which is neglected in this work.

Using the  number of e-folds, ${\rm d}N=H{\rm d}t$, as the  time variable and defining  the vectorial normalized white noise $\boldsymbol{\xi}$ via
\ba
\langle \boldsymbol{\xi} (N)\rangle =0 \,, \quad \quad 
\langle \xi_i (N) \xi_j (N')\rangle = \delta_{ij}~ \delta(N-N')  \,,
\ea
the above correlation functions allow us to express  the noises $(\boldsymbol{\tau},\boldsymbol{\sigma})$ in terms of the normalized white noise $\boldsymbol{\xi}$  as
\begin{align}
\boldsymbol{\sigma} (N)&\equiv
S_\nu(\varepsilon)~H^3~\boldsymbol{\xi}(N) \,,
\label{sigma-S}
\\
\boldsymbol{\tau} (N)&\equiv
T_\nu(\varepsilon)~H^4~\boldsymbol{\xi}(N) \,,
\label{tau-T}
\end{align}
in which the functions $S_\nu(\varepsilon)$ and $T_\nu(\varepsilon)$ are defined via 
\begin{align}
\label{S}
S_\nu(\varepsilon) &=
\dfrac{\Gamma(|\nu|) }{\pi} \dfrac{2^{-1+|\nu|}}{\sqrt{3\pi}}~\varepsilon^{{5 \over 2}-|\nu|} \,,
\\
T_\nu(\varepsilon) &=
\left\lbrace \begin{array}{lc}
S_\nu(\varepsilon)~\big| \dfrac{5}{2}-|\nu| \big|   &\nu \neq \pm 5/2 \\
\\
\dfrac{\varepsilon^2}{\sqrt{6}\pi} &\nu =\pm 5/2 \, .
\label{T}
\end{array}\right. 
\end{align}

The two coupled stochastic equations \eqref{X_long} and \eqref{Xdot_long} can be simplified further by considering the superhorizon behaviours of the auxiliary field $X$,
\begin{align}
X_\lambda (\eta,k) \propto~ {H^2}~\eta^{{5 \over 2}-|\nu|} \,,
\end{align}
where Eqs. \eqref{X_mode} and \eqref{Hanckel_limit} are used. Using the above approximation,  $\boldsymbol{\dot{\Pi}}^{\rm IR}$ is calculated to be 
\begin{eqnarray}
\boldsymbol{\dot{\Pi}}^{\rm IR}
&\simeq& \Big( |\nu|-{5 \over 2}-\big(|\nu|+{1 \over 2} \big)\epsilon_H \Big)H \boldsymbol{\Pi}_{\boldsymbol{}}^{\rm IR} \,.
\label{Pidot}
\end{eqnarray}
Now using Eq.~\eqref{Pidot} to eliminate $\boldsymbol{\dot{\Pi}}^{\rm IR}$ in favour of $H \boldsymbol{\Pi}^{\rm IR}$ in Eq.~\eqref{X_long} and combining the two coupled Eqs.~\eqref{X_long} and \eqref{Xdot_long} we  obtain the desired Langevin equation for the long mode $\boldsymbol{X}^{\rm IR}(N)$ as follows,  
\begin{eqnarray}
\dfrac{{\rm d}\boldsymbol{X}^{\rm IR}(N)}{{\rm d}N} &=& q_\nu \boldsymbol{X}^{\rm IR}(N) + \Big( S_\nu(\varepsilon)+\dfrac{T_\nu(\varepsilon)}{Q_\nu} \Big)
{H^2}~ \boldsymbol{\xi}(N) \,, \label{Xdot_long2}
\end{eqnarray}
where, to leading orders in slow-roll parameter, the parameters $q_\nu$ and $Q_\nu$ are defined via
\begin{eqnarray}
q_\nu &\equiv& \frac{1}{Q_\nu} \Big[ 
{(\nu-\dfrac{5}{2})(\nu+\dfrac{5}{2})-2(\nu^2-\dfrac{5}{4})\epsilon_H} \Big]  \, ,
\\
Q_\nu &\equiv&  |\nu|+{5 \over 2}-\big(|\nu|+{1 \over 2} \big)\epsilon_H
\,.
\end{eqnarray}

By defining the following dimensionless stochastic variable
\begin{eqnarray}
\boldsymbol{\cal X} = \dfrac{\boldsymbol{X}^{IR}}{X_{\rm ref}}
\,, \quad \quad  X_{\rm ref} &\equiv \sqrt{2\epsilon_H}M_P H \, ,
\label{calX2}
\end{eqnarray}
the Langevin equation~\eqref{Xdot_long2} can be cast into a dimensionless stochastic differential equation as follows, 
\begin{eqnarray}
{\rm d}\boldsymbol{\cal X}(N) &=& b_\nu~\boldsymbol{\cal X} ~{\rm d}N+ D_\nu(\varepsilon)~{\rm d}\boldsymbol{W}(N)  \,, \label{calX-Langevin-ap}
\end{eqnarray}
where $\textbf{W}$ is a three dimensional (3D) Wiener processes \cite{evans2013introduction} associated with the noises $\boldsymbol{\xi}$ via
\begin{eqnarray}
\mathrm{d}\boldsymbol{W}(N) &\equiv & \boldsymbol{\xi}(N) ~\mathrm{d}N \,,
\end{eqnarray}
while $b_\nu$ and $D_\nu$ represent the amplitude of the drift and the diffusion terms respectively,  given by
\begin{eqnarray}
b_\nu &\equiv& |\nu|-{5 \over 2}+\dfrac{75-28\nu^2+(10-8\nu^2)|\nu|}{2(5+2|\nu|)^2}~\epsilon_H 
\label{b}
\,,
\\
D_\nu(\varepsilon) &\equiv& 2\pi \sqrt{{\cal P}_\zeta} \Big( S_\nu(\varepsilon)+\dfrac{T_\nu(\varepsilon)}{Q_\nu} \Big)
\label{D}
\,,
\end{eqnarray} 
in which the power spectrum  ${\cal P}_\zeta$ is defined in Eq.~\eqref{Power-zeta}.

Eq.~\eqref{calX-Langevin-ap} is our master equation whose solutions were studied in the main text. 

It is useful to simplify the form of diffusion coefficient as a function of $\varepsilon$.  From Eqs. ~\eqref{S}, \eqref{T} and \eqref{D}, the diffusion coefficient up to leading order in $\varepsilon$ is given by
\begin{align}
\label{Dve}
D_\nu(\varepsilon) &=
\sqrt{6{\cal P}_\zeta} \times
\left\lbrace \begin{array}{lc}
\dfrac{2^{|\nu|}}{3} \, \dfrac{\Gamma(|\nu|)}{\sqrt{2\pi}} ~ \Big( 1 + \dfrac{\big| \dfrac{5}{2}-|\nu| \big|}{Q_\nu} \Big) \,\varepsilon^{{5 \over 2}-|\nu|} & |\nu|\neq 5/2
\\
1  &\nu = \pm 5/2 \, .
\end{array}\right.
\,
\end{align}
which will be used in the main text.

\bibliography{references} 
\bibliographystyle{JHEP}

\end{document}